\documentclass[fleqn,usenatbib]{mnras}
\usepackage{hyperref}
\usepackage{newtxtext,newtxmath}
\usepackage{bm}
\usepackage{amsmath}
\usepackage{gensymb}

\usepackage[T1]{fontenc}
\usepackage{ae,aecompl}

\usepackage{graphicx}	
\usepackage{amsmath}	
\usepackage{amssymb}	

\newcommand{\markus}[1]{{\color{black} #1}}

\title[Estimating z-distributions with logistic GPs]{Estimating redshift distributions using Hierarchical Logistic Gaussian processes}

\author[]{
Markus Michael Rau$^{1}$\thanks{E-mail: markusr@andrew.cmu.edu},
Simon Wilson$^{2}$,
Rachel Mandelbaum$^{1}$
\\
$^{1}$McWilliams Center for Cosmology, Department of Physics, Carnegie Mellon University, Pittsburgh, PA 15213\\
$^{2}$School of Computer Science and Statistics,
Lloyd Institute,
Trinity College,
Dublin,
Ireland\\
}

\date{Accepted XXX. Received YYY; in original form ZZZ}

\pubyear{2015}

\begin{document}
\label{firstpage}
\pagerange{\pageref{firstpage}--\pageref{lastpage}}
\maketitle

\begin{abstract}
This work uses hierarchical logistic Gaussian processes to infer true redshift distributions of samples of galaxies, through their cross-correlations with spatially overlapping spectroscopic samples. We demonstrate that this method can accurately estimate these redshift distributions in a fully Bayesian manner jointly with galaxy-dark matter bias models. We forecast how systematic biases in the redshift-dependent galaxy-dark matter bias model affect redshift inference. Using published galaxy-dark matter bias measurements from the Illustris simulation, we compare these systematic biases with the statistical error budget from a forecasted weak gravitational lensing measurement. If the redshift-dependent galaxy-dark matter bias model is mis-specified, redshift inference can be biased. This can propagate into relative biases in the weak lensing convergence power spectrum on the 10\% - 30\% level. We, therefore, showcase a methodology to detect these sources of error using Bayesian model selection techniques. Furthermore, we discuss the improvements that can be gained from incorporating prior information from Bayesian template fitting into the model, both in redshift prediction accuracy and in the detection of systematic modeling biases.
\end{abstract}




\section{Introduction}
In the era of large area photometric surveys like DES \citep[e.g.][]{2018ApJS..239...18A}, KiDS \citep[e.g.][]{2017MNRAS.465.1454H}, HSC \citep[e.g.][]{2018PASJ...70S...4A} and future surveys like LSST \citep[e.g.][]{2008arXiv0805.2366I}, WFIRST \citep[e.g.][]{2015arXiv150303757S} and Euclid \citep[e.g.][]{2011arXiv1110.3193L}, it becomes increasingly important to accurately model sources of systematic bias in weak gravitational lensing and large-scale structure probes \citep[e.g.][]{2018ARA&A..56..393M}. One of the most important systematics in the context of photometric surveys is associated with inaccurate measurements of distance, or redshift, using the galaxies' photometry \citep[e.g.][]{2006ApJ...636...21M, 2010MNRAS.401.1399B}. These photometric redshifts can be inferred by a variety of techniques. In `template fitting', we fit models for the spectral energy distribution (SED) of galaxies to their photometry \citep[e.g.][]{1999MNRAS.310..540A, 2000ApJ...536..571B,2006A&A...457..841I, 2006MNRAS.372..565F, 2015MNRAS.451.1848G, 2016MNRAS.460.4258L}. Alternative techniques use machine learning to infer a flexible mapping between the galaxies' photometry and redshift, using a spectroscopic training set \citep{2003LNCS.2859..226T, 2004PASP..116..345C, 2010ApJ...715..823G, 2013MNRAS.432.1483C,  2015MNRAS.449.1043B, 2015MNRAS.452.3710R, 2016A&C....16...34H}. More recently, a combination of these approaches has also been investigated \citep{2015arXiv151008073S, 2015MNRAS.450..305H}.

Traditionally, photometric redshifts are validated using redshifts derived from spectroscopic observations, which requires long exposure times. Since spectroscopic observations are consequently quite costly, the completeness of these surveys strongly depend on the galaxy type and it can be impractical to obtain complete spectroscopic validation samples that extend towards faint magnitudes \citep[see e.g.][]{10.1093/mnras/stu1424, 2015APh....63...81N}.

Cross-correlation techniques are a practical alternative to the aforementioned inference techniques that use information of the galaxies photometry  \citep[e.g.][]{2008ApJ...684...88N, 2013arXiv1303.4722M, 2013MNRAS.433.2857M, 2016MNRAS.462.1683S, 10.1093/mnras/stx691, Morrison2016, 2017arXiv171002517D, 2018MNRAS.477.1664G}. These methods use galaxy samples for which spectroscopic or highly accurate photometric redshifts are available, and cross correlate them with the sample {\color{black} for which redshifts need to be inferred. Since this sample typically has only photometric information available, we will refer to it as the `photometric sample', even though we are not using its photometry in the cross-correlation technique.} Except for the effect of cosmic magnification \citep[see e.g.][]{2005ApJ...633..589S}, this cross-correlation signal is only non-zero if both samples are at the same redshift. {\color{black} Therefore this methodology allows one to learn about the ensemble redshift distribution of the photometric sample. 
In contrast, the aforementioned techniques that use galaxy photometry constrain both the redshifts of the individual galaxies and the ensemble redshift distribution of the sample. In this paper we use the term `photo-z' distribution to denote ensemble redshift distributions of photometric samples of galaxies, i.e.\ not individual galaxies, even 
in the context of cross-correlation techniques  that do not use the photometry of the galaxies.  }

Extensions of this method employ different cosmological observables like weak gravitational lensing \citep[e.g.][]{2013MNRAS.431.1547B}, or jointly marginalize over cosmological parameters, while assuming Gaussian photometric selection functions \citep{2017MNRAS.466.3558M}. In more recent developments, \citet{2018arXiv180202581H} demonstrated that redshift information can be extracted even without the use of a spectroscopic reference sample solely from the correlation functions of the photometric sample itself{\color{black}, if the photo-z distributions follow a known family of distributions like Gaussians or simple mixtures thereof.} The first steps towards combining template fitting with cross-correlations have been made by \citet{2019MNRAS.483.2801S}, who use a simplified model for the redshift distribution and the cross-correlation measurement which fixes the galaxy-dark matter bias. They demonstrate nicely the improvements that can be gained by combining both complementary approaches in a Bayesian manner. Recently a similar approach has been applied to obtain redshift information for blended sources \citep{2019MNRAS.483.2487J}.

In this paper we extend and complement the aforementioned techniques by introducing the logistic Gaussian process as a flexible prior distribution to model photometric redshift distributions in a cross-correlation setting. This model easily allows the incorporation of prior knowledge of the redshift distribution and can be efficiently sampled with the methodology presented in this paper. As the redshift-dependent galaxy-dark matter bias of the photometric sample is degenerate with a flexible parametrization of the photometric redshift distribution, we marginalize over our uncertainty in this function. {\color{black} Gaussian processes are a popular machine learning method in cosmology that was used for example for machine learning-based photometric redshift estimation \citep[e.g.][]{2016MNRAS.462..726A} or to interpolate, and smooth, redshift histograms obtained using cross-correlation measurements \citep{2017MNRAS.465.4118J}. Our method differs from these previous applications of Gaussian processes, as we use the logistic Gaussian process as a prior to the shape of the logarithm of the redshift distribution and not as a regression model. In this way we can jointly marginalize over the parametrization of the photometric redshift distribution and other systematic uncertainties like galaxy-dark matter bias.

We specifically} discuss how systematic biases in the modelling of the redshift-dependent galaxy-dark matter bias affects redshift inference. Specifically we showcase how Bayesian model comparison can be used to detect these biases and investigate the effect of incorporating additional information from e.g.\ template fitting on both the predictive accuracy of our method and the robustness with respect to the aforementioned modelling systematics. \markus{We demonstrate this framework using a simulated mock data vector that uses published correlation function estimates from the Illustris simulation that provide galaxy-dark matter bias information. In this work we apply our methodology to two galaxy samples, i.e.\ the spectroscopic and photometric sample, that have different galaxy-dark matter bias properties. In the future we will apply our methodology to more realistic scenarios that accurately model the galaxy type selection of DESI-like spectroscopic surveys. }

This paper is structured as follows: in \S \ref{sec:methodology} we describe our methodology. This includes our modelling of the angular correlation functions, covariances, galaxy-dark matter bias and the design of our statistical model. In \S \ref{sec:results} we showcase the performance and accuracy of our methodology and study how systematic errors in the redshift dependent galaxy-dark matter bias impact redshift inference. We also propose Bayesian model comparison techniques as a way to detect these biases, which allows us to mitigate them by refining our modelling. We summarize and conclude in \S \ref{sec:summary_and_conclus}.

\section{Methodology}
\label{sec:methodology}
{\color{black} Measurements of statistics of the density field are  important for constraining cosmological parameters. Consider the product density $\rho(\mathbf{x_1}, \mathbf{x_2})$ of finding two galaxies with positions $\mathbf{x_1}$ and $\mathbf{x_2}$ in the infinitesimal volumes $\mathrm{d}V(\mathbf{x_1})$ and $\mathrm{d}V(\mathbf{x_2})$. If the point process that describes the galaxy density field at redshift $z$ is stationary and isotropic, this quantity only depends on the distance between the two galaxies $r = ||\mathbf{x_1} - \mathbf{x_2}||$ and we can write \citep[see e.g.][]{1999A&A...343..333K}
\begin{equation}
    \rho(\mathbf{x_1}, \mathbf{x_2}) = \overline{\rho}^2 \, (1 + \xi(r)) \, .
    \label{eq:def_density}
\end{equation}
Here, $\xi(r)$ denotes the two-point correlation function of galaxies\footnote{Note that the function $g(r) = 1 + \xi(r)$ is referred to as the `pair correlation function' in statistics \citep{stoyan1994fractals}.} and $\overline{\rho}$ denotes the average density of galaxies (per volume). The two-point correlation function is a statistic of the density field and can be used to constrain cosmological parameters. On the plane of the sky, we can define the angular correlation function $w(\theta)$ in analogy to $\xi(r)$ in Eq.~\eqref{eq:def_density} by replacing $\overline{\rho}$ by the average density of galaxies per unit area and $\xi(r)$ with the angular correlation function $w(\theta)$, where $\theta$ denotes the angular distance between the galaxy pair.

The basis of cross correlation redshift techniques are angular correlation functions between photometric and spectroscopic samples. The cross correlation amplitude\footnote{Here to be understood as the angular correlation function averaged over an angular interval according to Eq.~\eqref{eq:angular_averaging}.} $\widehat{w}_{\rm phot-spec}(z)$ between the photometric ($\rm phot$) and spectroscopic sample ($\rm spec$) at redshift $z$ can be written as
\begin{equation}
    \widehat{w}_{\rm phot-spec}(z) \propto p_{\rm phot}(z) \, p_{\rm spec}(z) \, b_{\rm phot}(z) \, b_{\rm spec}(z) \, w_{\rm DM}(z) \, .
    \label{eq:intro_cross_corr}
\end{equation}
The probability density functions $p_{\rm phot/spec}$ of the photometric and spectroscopic samples denote the probability of finding a galaxy in the sample at redshift $z$. These functions are therefore normalized to integrate to unity\footnote{The unit of the probability density function is the reciprocal unit of it's argument. While the redshift has no unit, we will later use the probability density function of the comoving distance of the galaxies. Then the unit of the probability density will be inverse distance.}. The terms $b_{\rm phot/spec}$ denote the galaxy-dark matter bias from the respective samples and $w_{\rm DM}(z)$ the amplitude of the dark matter density component. The squared galaxy-dark matter bias denotes the ratio between the two-point correlation function measured on the density field of tracers, i.e. the galaxies in the respective sample, and the two-point correlation function measured on the underlying dark-matter density field.

From Eq.~\eqref{eq:intro_cross_corr} we expect a significant angular cross correlation signal only between samples that overlap both spatially as well as in redshift. We then measure the cross-correlation between the photometric and spectroscopic samples binned in redshift intervals, e.g.\ tophat bins. In this way we can derive redshift information for the photometric sample from these cross-correlation signals. However as can be seen in Eq.~\eqref{eq:intro_cross_corr}, a redshift-dependent galaxy-dark matter bias evolution of the samples is degenerate with the photometric redshift distribution $p_{\rm phot}(z)$ that we want to infer. We therefore need to model these factors accurately.

The goal of this paper is to test our Bayesian redshift inference methodology specifically with respect to systematic errors in the modelling of a redshift-dependent galaxy-dark matter bias. Furthermore we investigate how these errors can be detected and corrected in a practical analysis. We note that our goal is not to forecast the redshift quality of any particular survey. We will make a number of simplifications in the modelling of the angular correlations function in \S \ref{subsec:modelling_angcorr}, the assumed redshift distribution and scale length model of the photometric and spectroscopic samples in \S \ref{subsec:modelling_galax_bias}, and the modelling of the covariance matrix in \S \ref{subsec:cov_matrix}. For easy reference we summarize these assumptions in \S \ref{subsec:summary}. The fiducial cosmological model used in this work is shown in Tab.~\ref{tab:cosmo_param}.
}

\subsection{Modelling the angular galaxy correlation function}
\label{subsec:modelling_angcorr}



We use the simple modelling of the angular correlation function used in previous cross-correlation analysis \citep{0004-637X-684-1-88, 2010ApJ...721..456M} that exploit the Limber approximation \citep{1953ApJ...117..134L} for the redshift projection and assume a power law shape of the correlation function at small scales. Using both approximations we are able to obtain an analytical solution to model the clustering signal. The gained computational efficiency allows us to more conveniently experiment with complex photometric redshift parametrizations, while still maintaining a practical methodology within the testable and well known limits of the imposed approximations.

For small scales below $\sim 15~{\rm Mpc/h}$ \citep{2002ApJ...571..172Z, 2007A&A...473..711S, 2018MNRAS.475..676S}, but above $\sim 1~{\rm Mpc/h}$ \citep[][Fig. 1]{2018MNRAS.475..676S} we can approximate the correlation function $\xi(r)$ as a power law
\begin{equation}
    \xi(r) = \left(\frac{r}{r_0}\right)^{-\gamma} \, ,
    \label{eq:two_point_corrfnkt}
\end{equation}
where the clustering scale length $r_0$ and exponent $\gamma$ are a function of redshift $z$, or equivalently comoving distance. As explained in the next section, we will model the redshift dependence of these values using the measured functions from the Illustris simulation, for different galaxy samples, from \citet{2018MNRAS.475..676S}.
The Limber approximation assumes that the redshift distribution varies little compared with the coherence length of the considered clustering density field \citep[][p. 43]{2001PhR...340..291B}. Following \citet{2007A&A...473..711S} we write the angular correlation function as
\begin{equation}
    w(\theta) = \int_{0}^{\infty} d\overline{r} p_1(\overline{r}) p_2(\overline{r}) \int d\Delta r \, \xi(R, \overline{r}) \, ,
\end{equation}
where we used the variables $R = \sqrt{\overline{r}^2 \theta^2 + \Delta r^2}$ and $\overline{r} = \frac{r_1 + r_2}{2}$ and $\Delta r = r_2 - r_1$ and denote with $r_1$ and $r_2$ the radial distances of a pair of galaxies. {\color{black} The functions $p_{1/2}(\overline{r})$ denote the sample comoving distance distributions that are connected with the redshift distributions $p(z)$ as $p(\overline{r}) = p(z) \left|dz\large/d\overline{r}\right|$.}

Using the power law approximation in Eq.~\eqref{eq:two_point_corrfnkt}, the angular correlation function can be expressed as
\begin{equation}
    w(\theta) = \int_{0}^{\infty} \, d\overline{r} \, A_w(\overline{r}) \, \left(\frac{\theta}{{\rm 1 RAD}}\right)^{1 - \gamma(\overline{r})} \, ,
\end{equation}
where
\begin{equation}
    A_w(\overline{r}) = \sqrt{\pi} \,  r_0(\overline{r})^{\gamma(\overline{r})} \, \left(\frac{\Gamma(\gamma(\overline{r})/2 - 1/2)}{\Gamma(\gamma(\overline{r})/2)}\right) \,  p_1(\overline{r}) \,  p_2(\overline{r}) \, d_A(\overline{r})^{1 - \gamma(\overline{r})} \, .
\end{equation}
Here, $d_A(\overline{r})$ denotes the angular diameter distance, $\Gamma(x)$ are gamma functions.

To simplify the calculation, we discretize the comoving distance-dependence of the scale length $r_0(\overline{r})$, the exponent $\gamma(\overline{r})$ and the photometric redshift distribution into the same $N$ comoving distance bins
\begin{equation}
    w(\theta) = \sum_{i = 1}^{N} \overline{A}_{w}^{i} \left(\frac{\theta}{\rm 1 \, RAD}\right)^{1 - \gamma_i} \, ,
\end{equation}
where
\begin{equation}
    \overline{A}_{w}^{i} = \sqrt{\pi} \,  r_{0, i}^{\gamma_i} \, \left(\frac{\Gamma(\gamma_i/2 - 1/2)}{\Gamma(\gamma_i/2)}\right) \,  p_1^{i} \,  p_2^{i} \, \int_{r_{\rm L, i}}^{r_{\rm R, i}}d_A(\overline{r})^{1 - \gamma_i} d\,\overline{r} \, .
\end{equation}
Here $r_{\rm L, i}$ and $r_{\rm R, i}$ denote the left and right edges of comoving distance bin $i$.

As discussed in more detail in the next section, we bin the angular correlation function in $\theta$ space \citep[see e.g.][]{2017MNRAS.468.2938S}
\begin{equation}
    \widehat{w} = \frac{1}{\Delta \Omega} \int \, d\Omega \, w(\theta) \, ,
    \label{eq:angular_averaging}
\end{equation}
where $\widehat{w}$ denotes the binned angular correlation function, and $\Omega$ denotes the solid angle, where
\begin{equation}
    \Delta \Omega = 2 \pi \int_{\theta_L}^{\theta_R} d\theta^{'} \sin{\theta^{'}} \approx  \pi \, (\theta_R^2 - \theta_L^2) \, .
    \label{eq:solid_ang_norm}
\end{equation}
Here we used the small angle approximation $\sin(\theta) \sim \theta$, which is valid to good accuracy within small angular intervals $\theta \in [\theta_L, \theta_R]$. We thus obtain
\begin{equation}
    \widehat{w} = \frac{2}{\theta_R^2 - \theta_L^2} \sum_{i = 1}^{N} \, \overline{A}_{w}^{i} \left(\frac{\theta_{R}^{3 - \gamma_i} - \theta_{L}^{3 - \gamma_i}}{3 - \gamma_i}\right)
    \label{eq:w_theta_binned}
\end{equation}

To model the scale length of the photometric and spectroscopic samples, we assume linear biasing \citep[see e.g.][]{2010ApJ...721..456M}:
\begin{equation}
    r_{\rm 0, sp}(\overline{r})^{\gamma_{\rm sp}(\overline{r})} = \sqrt{r_{\rm 0, ss}(\overline{r})^{\gamma_{\rm ss}(\overline{r})} r_{\rm 0, pp}(\overline{r})^{\gamma_{\rm pp}(\overline{r})}} \, ,
    \label{eq:linear_biasing}
\end{equation}
where $r_{\rm 0, sp}(\overline{r})$, $r_{\rm 0, ss}(\overline{r})$, $r_{\rm 0, pp}(\overline{r})$ and $\gamma_{\rm sp}(\overline{r})$, $\gamma_{\rm ss}(\overline{r})$, $\gamma_{\rm pp}(\overline{r})$ denote the comoving distance-dependent scale length and slope of the two point correlation function (Eq.~\ref{eq:two_point_corrfnkt}) for the cross-correlation between the spectroscopic and photometric samples, the spectroscopic sample and the photometric sample.

Linear biasing is also assumed to obtain the redshift-dependent galaxy bias models for the modelling of the covariance matrix in \S \ref{subsec:cov_matrix}, that employ angular correlation power spectra of the cross- and auto-correlations
\begin{equation}
    b_{\rm sp, ss, pp}(\overline{r})^2 = \left(\frac{r_{\rm 0, sp, ss, pp}(\overline{r})}{r_{\rm 0, tot}(\overline{r})}\right)^{\gamma(\overline{r})} \, .
    \label{eq:bias_conversion}
\end{equation}
{\color{black} Here $r_{\rm 0, tot}(\overline{r})$ denotes the scale length of the total matter component measured in \citet{2018MNRAS.475..676S} on scales $5 h^{-1}{\rm Mpc} - 20 h^{-1}{\rm Mpc}$. In this range, the total matter power spectrum is very similar to the dark-matter only power spectrum \citep[][Fig. 9]{2018MNRAS.475..676S}. We therefore use it for the definition of the galaxy-dark matter bias in Eq.~\eqref{eq:bias_conversion}. Within the considered scales, we expect that the influence of scale-dependent biasing is quite low \citep[][Fig. 19, top right panel]{2018MNRAS.475..676S}. We therefore use the same redshift-dependent power law slope $\gamma$ for the dark matter and galaxy components.}

Similarly, since the redshift-dependent power law slope $\gamma$ does not depend much on stellar mass, as found in \citet[][Fig. 14]{2018MNRAS.475..676S}, we will use the same power spectrum slope for all galaxy samples considered in this forecast. The used model corresponds to the stellar mass sample $8.5 < \log{(M_{*}[h^{-2} M_{\odot}])} < 9.0$ shown in \citet[][Fig. 14]{2018MNRAS.475..676S}. Furthermore we will assume that the slope and scale length of the two point correlation function $\gamma$ and $r_{\rm 0, ss}$ can be measured sufficiently accurately in the spectroscopic sample, compared with the uncertainty in the redshift-dependent scale length of the photometric sample. We will therefore fix these values in the statistical analysis described in \S \ref{subsec:stats_modelling}, based on the high signal-to-noise ratio expected for next generation spectroscopic surveys \citep[see, e.g.,][]{2016arXiv161100036D}. {\color{black} This is similar to the approach used in traditional cross-correlation redshift estimates, where one fits the scale length and power-law slope of the spectroscopic sample and subsequently uses these fitted parameters in the cross-correlation analysis assuming linear biasing \citep[e.g.][]{2012ApJ...745..180M}. }

\subsection{Modelling the redshift distribution and scale length of the galaxy samples}
\label{subsec:modelling_galax_bias}

\begin{figure*}
\begin{center}
\includegraphics[width=\linewidth]{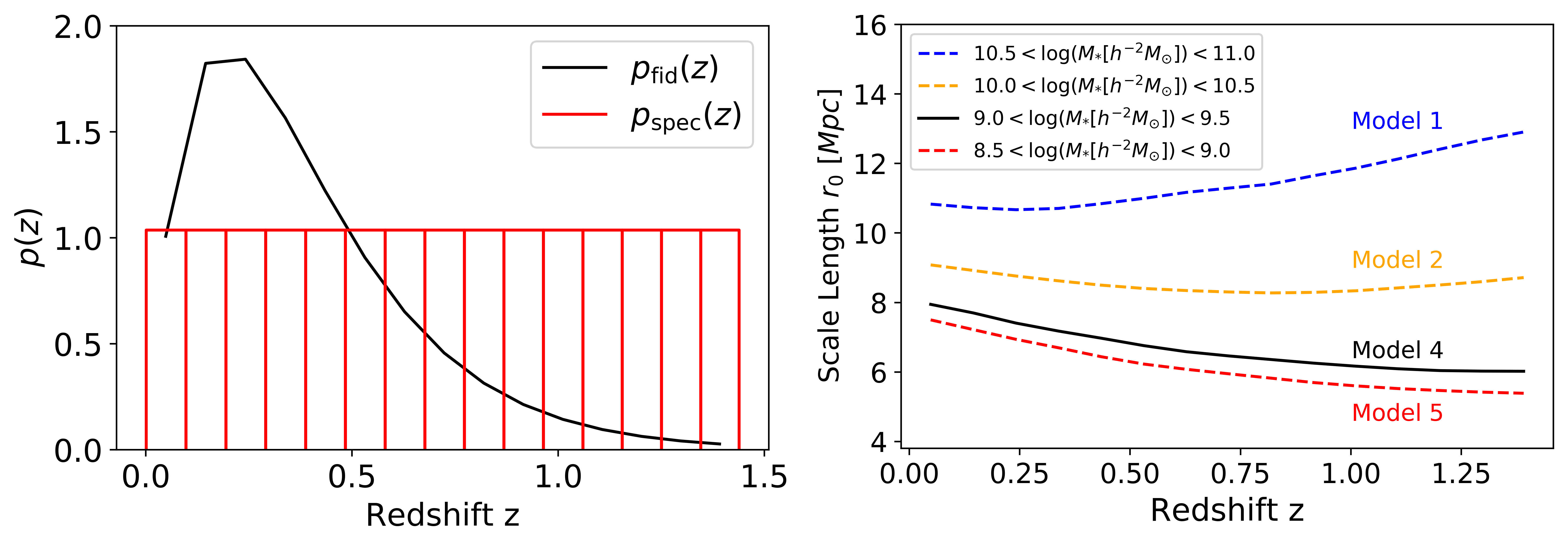}
\caption{\label{lab:galaxy_dm_bias} \textit{Left:} Redshift distribution of the background sample and spectroscopic tophat functions. \textit{Right:} Evolution of the scale length as a function of redshift. We omit a `Model 3' here that would correspond to the stellar mass interval between Model 2 and Model 4 and can be found in \citet[][Fig. 13]{2018MNRAS.475..676S}. We do not use this stellar mass bin here, but still want to indicate with the numbering that there is a galaxy population between those shown here. }
\end{center}
\end{figure*}
\begin{figure*}
\begin{center}
\includegraphics[width=\linewidth]{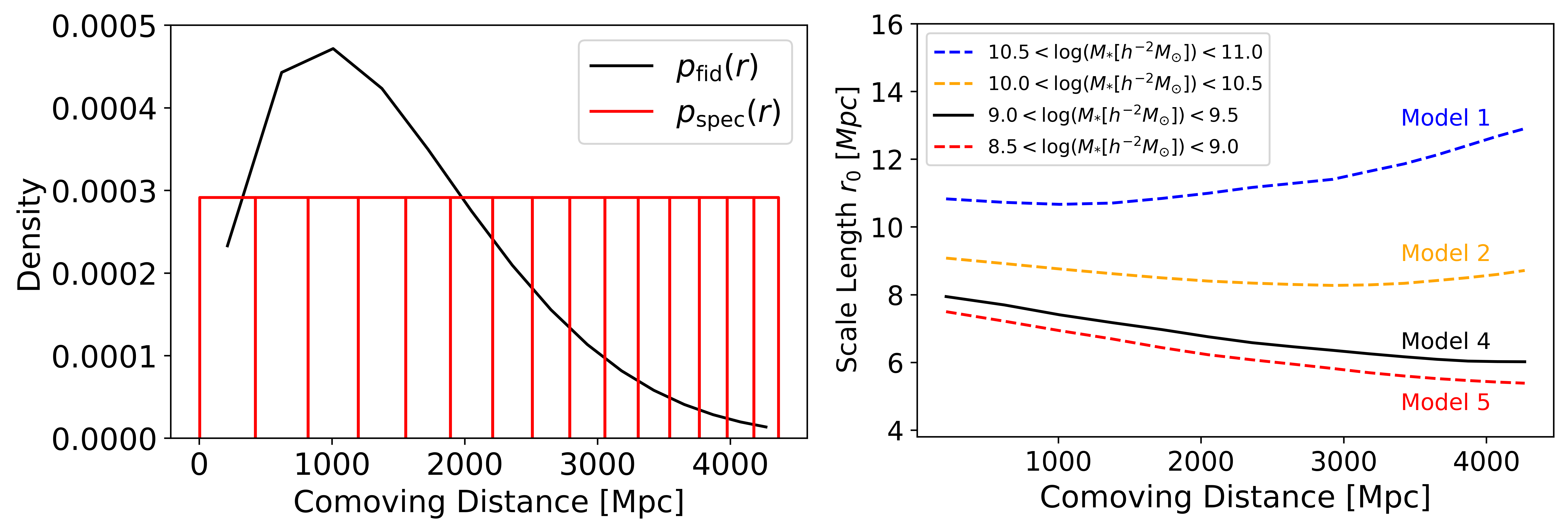}
\caption{\label{lab:galaxy_dm_bias_comov} \textit{Left:} Distribution of comoving distance of the background sample and spectroscopic tophat functions. \textit{Right:} Evolution of the scale length as a function of comoving distance. }
\end{center}
\end{figure*}
The left panels of Fig.~\ref{lab:galaxy_dm_bias} and Fig.~\ref{lab:galaxy_dm_bias_comov} show the assumed redshift and comoving distance distributions. The black line shows the redshift distribution of the photometric sample and the red bins those of the spectroscopic sample. The right panels of Fig.~\ref{lab:galaxy_dm_bias} and Fig.~\ref{lab:galaxy_dm_bias_comov} plot the scale length models of our mock galaxies.  {\color{black} The redshift distribution covers a relatively small redshift range up to $z < 1.5$ that is designed to ensure that the assumptions we impose on the redshift evolution of the galaxy-dark matter bias and the power law exponent can be expected to be valid to a good degree. We refer to the following sections \S \ref{subsec:generating_mock_data} and \S \ref{subsec:summary} for a more detailed discussion of these assumptions. We picked a tailed distribution to ensure that our methodology is tested to be robust against cross-correlation measurements with very different signal-to-noise levels. We note however that the particular shape of the photometric redshift distribution does not influence our results on a fundamental level, as we parametrize the photometric redshift distribution using a logistic Gaussian process on the (log) histogram bins, which is a very flexible non-parametric approach.

It has to be noted however, that the redshift binning of the spectroscopic sample can introduce an undesired smoothing effect, if these bins are chosen to be too large. For more peaked distributions, we therefore might have to consider a finer binning. This effect has been studied in \citet{2017MNRAS.466.2927R}, where recommendations about bin width selection, as well as on methods to detect and correct the oversmoothing effect are described. Since our distribution is quite smooth, we expect that our binning scheme is fine enough to capture the structure of the distribution well.  }

The redshift evolution of the scale length models has been extracted from \citet[][Fig. 13]{2018MNRAS.475..676S}\footnote{We extract these functions directly from the paper plots using the WebPlotDigitizer software \url{https://automeris.io/WebPlotDigitizer/}.}. The redshift evolution of the exponent $\gamma$ is less dependent on stellar mass as discussed in \citet{2018MNRAS.475..676S}. For simplicity we therefore choose a common function for these samples that corresponds to the stellar mass bin of Model 5 in Fig.~\ref{lab:galaxy_dm_bias}, i.e. $8.5 < \log{(M_{*}[h^{-2} M_{\odot}])} < 9.0$. The different scale length models correspond to the stellar mass ranges of the different galaxy samples that we want to consider in this work. In the following we will denote these models with the numbers quoted in the plot. Since we will compare different combinations of galaxy-dark matter bias models for the spectroscopic and photometric samples, we will use different scale length factor models for each sample. As the redshift, or comoving distance, evolution of the scale factor cannot be measured directly for the photometric sample, we need to parametrize our uncertainty in this quantity. \citet{2012ApJ...745..180M} use a constant factor to correct the scale length models from the measured spectroscopic sample to match the photometric one, i.e. $r_{\rm 0, spec}(z) \propto r_{\rm 0, phot}(z)$. However we found that a constant shift $\Delta$, i.e. $r_{\rm 0, spec}(z) = \Delta +  r_{\rm 0, phot}(z)$ provides a better fit to offset the spectroscopic scale length model (Model 5) to the assumed photometric scale factor model (Model 4). {\color{black} Furthermore we would like to study more complicated cases, where the difference in the photometric/spectroscopic scale length models is not a simple offset, but depends in a more complicated manner on the comoving distance.} This motivated us to parametrize the comoving distance-dependence of the scale length as Chebychev polynomials
\begin{equation}
    r_{\rm 0, pp}(\overline{r}) = \sum_{i=0}^{M} \Delta_i T_i(\overline{r}) \,
\end{equation}
where the $T_i(\overline{r})$ are given by the recurrence relation
{\color{black}
\begin{align}
    T_0(\overline{r}) &= 1 \\
    T_1(\overline{r}) &= \overline{r} \\
    T_{n + 1}(\overline{r}) &= 2 \, \overline{r} \, T_n(\overline{r}) - T_{n - 1}(\overline{r}) \, .
\end{align}
If the coefficients of the higher order terms $T_i$ for $i>0$ are the same for the spectroscopic and photometric samples, this parametrization is equivalent to the previously discussed fit between Model 5 and 4. }

We found that the expansion in Chebychev polynomials converged rapidly to the true function. By the properties of this basis, omitting higher order terms in this expansion leads to an intrinsic sparsity in the approximation and they provide good fits to the scale length for $M=2$. The constant offset $\Delta$ then corresponds to the $\Delta_0$ parameter in this expansion. We note that we use the Chebychev expansion {\color{black} only} as a way to interpolate the comoving distance-dependence of the scale length. It is no substitute for a physically-motivated galaxy-dark matter bias model.

\subsection{Modelling the Covariance Matrix}
\label{subsec:cov_matrix}
{\color{black} The following section describes the modelling of the theoretical covariance matrix. The previous section used a simplified modelling of the angular correlation function. Since computational speed and modelling simplicity is not an advantage for the modelling of the covariance matrix, we use a more accurate approach here.}
We model the covariance matrix of the angular correlation functions binned in angular bins $i$, $j$ for the redshift bins $a$ and $b$ by transforming the covariance matrix of the corresponding angular correlation power spectra for redshift bins $a$ and $b$ as \citep[see e.g.][]{2011MNRAS.414..329C, 2017MNRAS.468.2938S}
\begin{equation}
    \Sigma_{(a, b), (i, j)} = \sum_{l \geq 2} \left(\frac{2 \ell + 1}{4 \pi}\right)^2 \left[\widehat{L}_{\ell}^{i} \, \widehat{L}^{j} \, \Sigma_{l, l^{'}}^{(a,b)}\right] \, ,
    \label{eq:correlation_theta}
\end{equation}
where $\Sigma_{l, l^{'}}^{a, b}$ denotes the covariance matrix of the angular correlation power spectra for a pair of redshift bins $a$ and $b$ and
\begin{align}
    \begin{split}
    \widehat{L}_{\ell}^{i}
    &=  \frac{2 \pi}{\Delta \Omega_i} \frac{1}{2 \ell + 1} \, \tilde{L} \\
    \tilde{L} &= L_{\ell - 1}\left(\cos{(\theta_R^i)}\right) - L_{\ell + 1} \left(\cos{(\theta_R^i )}\right) \\ &- L_{\ell - 1} \left(\cos{(\theta_L^i)}\right) + L_{\ell + 1}\left(\cos{(\theta_L^i)}\right) \, ,
     \end{split}
\end{align}
are Legendre polynomials $L_{\ell}\left(x\right)$ binned in the respective angular bins $i$ or $j$. Here $\theta_L^{i}$ and $\theta_R^{i}$ denote the left and right edges of the angular bins and $\Delta \Omega_i$ the solid angle integral defined in Eq.~\eqref{eq:solid_ang_norm}.
\begin{figure*}
\begin{center}
\includegraphics[width=\linewidth]{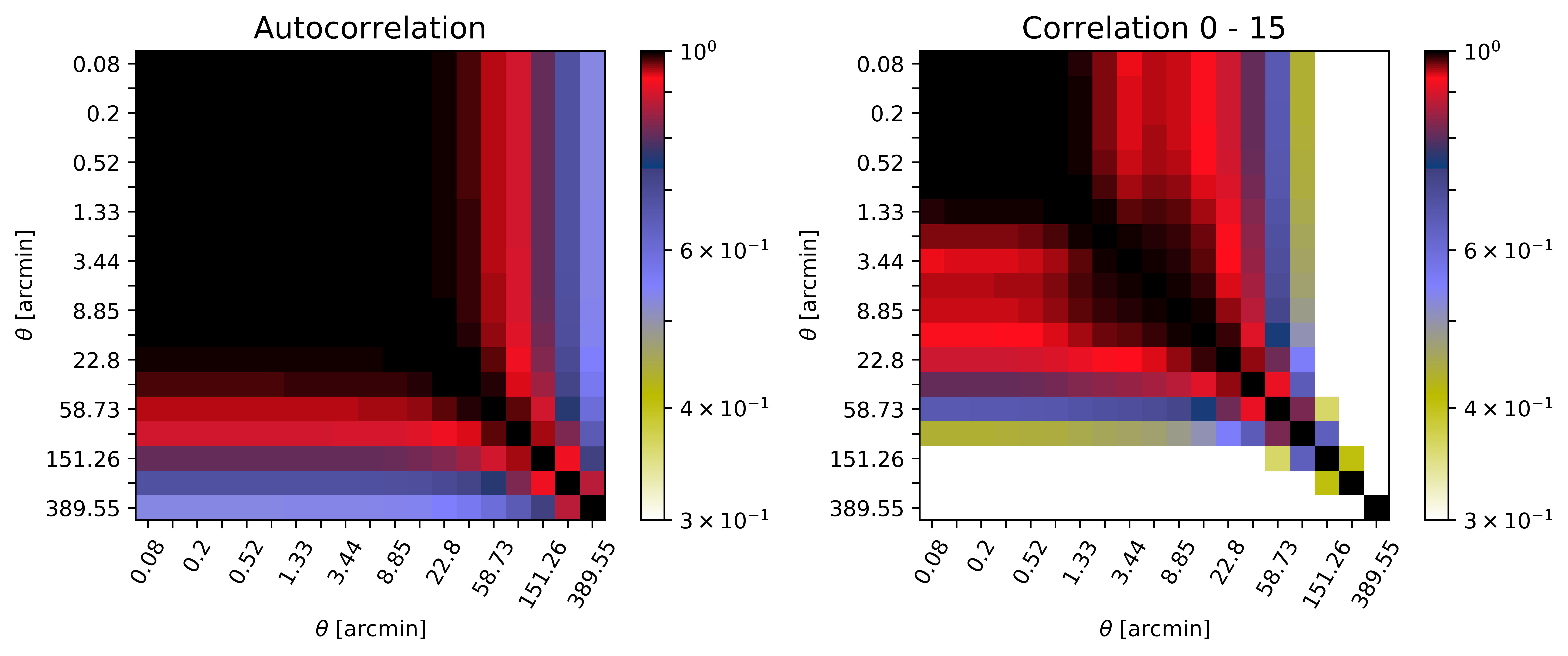}
\caption{\label{fig:corr_matrix}  Correlation matrix of angular correlation functions for the auto correlation of the photometric redshift distribution (\textit{Left}) and cross-correlation between the photometric redshift distribution and the last spectroscopic tophat bin (\textit{Right}). The bin sizes {\color{black} and colorbar} are spaced logarithmically. {\color{black} We set a lower limit on the correlation coefficients $|\rho_{i, j}| < 0.3$ in this plot for better visibility of structure and consistency between the two panels. Accordingly, angular bins with $\theta > 1^{\circ}$ in the right panel are only weakly correlated with sub-degree scales with a correlation coefficient $|\rho_{i, j}| < 0.3$. } }
\end{center}
\end{figure*}
We model the covariance matrix between two angular correlation power spectra, that correspond to a pair of redshift bins $(i, j)$ and $(k, l)$ as
\begin{equation}
	\Sigma_{(i, j)}^{(k, l)}(\ell) = A(\ell) \, \left(\bar{C}^{(i, k)}(\ell) \, \bar{C}^{(j, l)}(\ell) + \bar{C}^{(i, l)}(\ell) \, \bar{C}^{(j, k)}(\ell)\right) \, \, ,
	\label{eq:cov_matrix_ell}
\end{equation}
{\color{black} which neglects correlations between neighboring angular modes $\ell$ and non-Gaussian contributions.}
The cosmic variance factor
\begin{equation}
A(\ell) = \frac{\delta_{\ell, \ell^{'}}}{(2 \ell + 1) f_{\rm sky}}
\end{equation}
is inversely proportional to the fractional sky coverage $f_{\rm sky}$ and $\bar{C}^{(i, j)}(\ell)$ denotes the shot noise contribution to the angular power spectra
\begin{equation}
	\bar{C}^{(i, j)}(\ell) = C^{(i, j)}(\ell) + \frac{\delta_{i, j}}{\bar{n}_{g}^{i}} \, .
\end{equation}
The shot noise $\bar{n}_{g}^{i}$ denotes the number of galaxies per steradian in the respective sample.
The correlation power spectrum of a cosmological density field, e.g.\ the density field of galaxy positions or the corresponding field of galaxy shapes in lensing, can be defined as \citep[e.g.][]{2015MNRAS.451.4424K}
\begin{equation}
    C_{\ell}^{i, j} = \frac{2}{\pi} \int W^i(\ell, k) W^{i}(\ell, k) k^2 P(k) dk \,
\end{equation}
where $P(k)$ denotes the matter power spectrum and $W^i(\ell, k)$ are weighting functions that `project' the matter power spectrum along the redshift dimension. For galaxy clustering the weighting function is defined as
\begin{equation}
    W_{\rm clus}(\ell, k) = \int b_g(k ,z) n(z) j_{\ell}(k r(z)) D(z) dz \,
\end{equation}
where $b_g(k, z)$ denote the (potentially) redshift and scale dependent galaxy-dark matter bias, $n(z)$ denotes the photometric redshift distribution, $j_{\ell}(k r(z))$ denote the spherical bessel functions and $D(z)$ the growth function.
For weak gravitational lensing the weighting function takes the form
\begin{equation}
    W_{\rm lens}(\ell, k) = \int q(z) j_{\ell}(k r(z)) D(z) dz \, ,
\end{equation}
where $q(z)$ denotes the lensing weight
\begin{equation}
    q(z) = \frac{3 H_0^2 \Omega_m}{2 c^2} \frac{r(z)}{a(z)} \int_{r_{\rm hor}}^{r} dr^{'} n(r(z^{'})) \left(\frac{r(z^{'}) - r(z)}{r(z^{'})}\right) \, ,
\end{equation}
where $a(z)$ is the scale factor evaluated at redshift $z$ and $r_{\rm hor}$ is the comoving horizon.

The cosmological parameter values and forecast assumptions made in this work are listed in Tab.~\ref{tab:cosmo_param} {\color{black} and the assumed redshift distributions are shown in Fig. \ref{lab:galaxy_dm_bias}. The assumptions on number density and sky coverage are selected in analogy to \citet{2015MNRAS.451.4424K} and would correspond to a DES Y5-like survey. However the redshift distribution covers a smaller redshift range, as discussed in the previous section. This implies that our signal-to-noise will likely be overestimated, which will make systematic biases due to a misspecified redshift dependent galaxy-dark matter model more significant, compared with the statistical error budget. We reiterate that we do not intend to forecast the performance of a particular survey, but rather to test the robustness of our redshift inference to a redshift-dependent galaxy-dark matter bias mis-specification. }  We use the cosmosis\footnote{\url{https://bitbucket.org/joezuntz/cosmosis/wiki/Home}} \citep{2015A&C....12...45Z} software to estimate the non-linear matter power spectrum for these parameters and the `LimberJack' code \footnote{\url{https://github.com/damonge/LimberJack}} to calculate the galaxy angular power spectra, where we use the Limber approximation for $\ell > 60$ and the exact calculation for smaller modes \citep[see e.g.][\S 4]{2011MNRAS.412.1669T}. We perform the summation in Eq.~\eqref{eq:correlation_theta} up to $\ell = 5000$ to ensure convergence. {\color{black} We note that the covariance matrix of the angular correlation power spectra exhibits correlations between the auto-correlation of the photometric bin and the cross-correlations with the spectroscopic tophat bins. These correlations are largest for low angular modes and decrease for larger modes. As these cross-correlations are typically ignored in similar cross-correlation analyses, where the auto-correlation and the cross-correlations are fit individually \citep[e.g.][]{0004-637X-684-1-88, 2010ApJ...721..456M}, we will only consider the diagonal component of the covariance matrix for simplicity. However we do stress that these cross terms have been used in previous applications of the cross-correlation technique \citep[e.g.][]{2012ApJ...745..180M} and should be included in a practical application to data. }

The left panel of Fig.~\ref{fig:corr_matrix} shows the correlation matrix of the angular correlation function for the example of the auto-correlation of the photometric sample. Similarly the right panel of Fig.~\ref{fig:corr_matrix} shows the correlation matrix of the angular correlation function for the cross-correlation between the photometric sample and the last spectroscopic tophat bin. We see in both panels, that the correlation of sub-degree angular scales is very high and decreases for correlations on larger angular scales. We note however that this result strongly depends on the impact of shot noise on the covariance, and therefore on the considered galaxy sample. {\color{black} Note that in order to better resolve the structure of the correlation matrix on small scales, we decrease the color range by setting a lower limit on the correlation coefficients in the plot to $|\rho_{i, j}| > 0.3$. }

Due to the high correlation on sub-degree scales, we consider a single angular bin from $\theta \in [0.1, 1.0] \, {\rm arcmin}$, which is in the regime where the Limber approximation (see \S \ref{subsec:modelling_angcorr}) is quite accurate. These scales are also comparable to the angular bins within $0.06 - 6 \, [\rm{arcmin}]$ that have been used in \citet{2010ApJ...721..456M}. {\color{black} We note that this excludes larger scales that could be used to constrain cosmological parameters in a subsequent analysis. However the significant covariance between these larger (degree) scales and the sub-degree regime, {\color{black} particularly in the autocorrelation} implies that we need to account for this {\color{black} covariance}. We will further comment on this point in \S \ref{subsec:summary}.  }
\begin{table}
\centering
\caption{Cosmological parameter values in analogy to \citet{2018MNRAS.475..676S} and forecast assumptions for a photometric and spectroscopic survey similar to \citet{2015MNRAS.451.4424K}.}
\label{tab:cosmo_param}
\begin{tabular}{l  l  l  l  l  l }
 $\Omega_m$ & 0.3089 & $h$ & 0.9774 & $f_{\rm sky}$ & 0.12 \\\hline
 $\Omega_b$ & 0.0486 &  $\sigma_8$ & 0.6774 & $n_g^{\rm phot} \, [{\rm arcmin}^{-2}]$ & 10      \\\hline
 $\Omega_\Lambda$ & 0.6911 & $n_s$ & 0.9667 & $n_g^{\rm spec} \, [{\rm arcmin}^{-2}]$ & 0.56  \\\hline
\end{tabular}
\end{table}
\markus{
\subsection{Generating the mock data vector}
\label{subsec:generating_mock_data}
As discussed in the previous section, we find a large correlation between angular correlation function bins on small angular scales. Similar to \citet{2013arXiv1303.4722M} we therefore generate our mock data for the photometric galaxy angular auto- and cross-correlations with the spectroscopic tophat bins using a single angular bin within $[0.1, 1.0] \, \rm{arcmin}$ following Eq.~\eqref{eq:w_theta_binned}.

This correlation function is generated assuming a redshift-dependent scale length and power spectrum exponent $\gamma(r)$ model that we discretize within the comoving distance bins shown in Fig.~\ref{lab:galaxy_dm_bias} as described in \S \ref{subsec:modelling_angcorr}. We note that the binning of the redshift distribution and the discretization of $r_0(r)$ and $\gamma(r)$ are performed in comoving distance. We reiterate that the creation of the covariance matrix uses a more accurate modelling of the correlation functions. Since the software we use to obtain galaxy angular correlation power spectra in \S~\ref{subsec:cov_matrix} assumes a redshift distribution, we transform the histograms defined as comoving distance distributions $p(r)$ into redshift space $p(z)$ according to $p(z) = p(r) |dr/dz|$. Since the histogram assumes a uniform distribution in comoving distance, this transformation leads to tilted histogram bins and tophat functions. Since this is an artifact of the histogram discretization, we re-bin at the histogram midpoints for plots like Fig.~\ref{lab:galaxy_dm_bias}. However, to ensure consistency with the modelling of the angular correlation function data vector, we transform all comoving distance tophat functions and histogram bins into redshift space without re-binning before the angular correlation power spectra are obtained.

We reiterate that the scale length model strongly varies as a function of stellar mass, while the model for $\gamma(z)$ is quite similar for different galaxy populations within the redshift range ($z < 1.5$) considered in this work. Thus, while we will use different scale length models in our mock data vector, we will choose the same model for $\gamma(r)$ that corresponds to Model 5 in Fig.~\ref{lab:galaxy_dm_bias}.

We note that the long-tailed shape of the photometric redshift distribution shown in the left panel of Fig.~\ref{lab:galaxy_dm_bias} ensures that the sampling tested in the following sections has to prove its robustness against significant variations in signal-to-noise in the parameters that describe the photo-z distribution. As this work does not consider inhomogeneous spectroscopic samples, i.e.\ spectroscopic samples that consist of very different galaxy populations across redshift, we do not extend our redshift range beyond $z>1.5$, where a realistic spectroscopic sample will consist of high-redshift quasar samples with quite different clustering properties compared with the rest of the sample.

The modelling of the data vector also uses Eq.~\eqref{eq:w_theta_binned}, however the scale length offset $\Delta_0$ (see \S~\ref{subsec:modelling_galax_bias}) is treated as a random variable. In \S~\ref{subsec:impact_wrong_scale_length} we study the effect of biased scale length models on the accuracy of the reconstructed photo-z distribution.
}
{\color{black}
\subsection{Summary of modelling assumptions}
\label{subsec:summary}
\begin{table*}
\centering
\caption{Summary of modelling assumptions in the generation of the data vector, modelling of the mock data vector and modelling of the covariance matrix.\label{tab:modelling_assumptions} }
\label{tab:cosmo_param}
\begin{tabular}{l  l  l l}
  Assumption  & Mock Data Vector & Data Vector Model & Covariance matrix  \\\hline\hline
  Limber approximation & Yes  & Yes & No (Exact modelling)  \\
  Power law correlation function & Yes & Yes & No (Exact modelling) \\
  Linear Biasing & Yes & Yes & Yes \\
  Scale-dependent bias & No & No & No \\
  Redshift-dependent bias & Yes & Yes & Yes \\
  Covariance between $\ell$  & N/A & N/A & No \\
  Covariance auto- and cross-correlations & N/A & N/A & No \\
  Cosmic Variance & N/A & N/A & Yes\\
  Account for cosmological parameter dependence & N/A  & No & N/A \\
  Simplified treatment of spectroscopic survey & Yes & Yes & Yes \\
\end{tabular}
\end{table*}
In the following we summarize the modelling assumptions made when generating the mock data vector, the modelling of this data vector and the modelling of the covariance matrix. These points are also listed in Tab.~\ref{tab:modelling_assumptions} for easy reference. We list the assumption in the first column and list if this affects the generation of the mock data vector, the modelling of the data vector and the modelling of the covariance matrix in the subsequent columns.
\paragraph*{Limber Approximation and Power Law correlation function}
We use the Limber approximation to model the angular correlation functions within a single angular bin of $\theta \in [0.1, 1.0] \, {\rm arcmin}$. The Limber approximation in combination with a power-law correlation function is the basis of many cross correlation methods \citep[e.g.][]{0004-637X-684-1-88, 2010ApJ...721..456M, 2013arXiv1303.4722M}. We use this assumption in both the generation and modelling of the data vector. This analytical model allows us to quickly evaluate the correlation functions, which is very beneficial for efficient sampling. The Limber approximation can be expected to be accurate on the percent level within the considered scales \citep[see e.g.][]{2007A&A...473..711S}. We expect that deviations from the power-law correlation function can introduce a larger source of modelling bias, especially on smaller scales and high redshift $z>1.5$, where scale-dependent galaxy-dark matter bias will become important. For the modelling of the covariance matrix, we do not assume a power law correlation function and do not impose the Limber approximation. Since the advantages of the Limber approximation, i.e.\ convenient modelling and computational speed, do not affect the generation of the covariance matrix, we 
use the more exact treatment here.
\paragraph*{Galaxy-Dark Matter bias}
While we investigate the effect of a redshift-dependent galaxy-dark matter bias, we do not address the issue of a possible scale dependence. We neglect the stellar mass dependence of the correlation exponent $\gamma$ and use the same model for all galaxy samples. Furthermore we assume linear biasing throughout the paper. These assumptions are made for the creation of the mock data vector and its modelling.

Since we use the full modelling of the power spectrum for the modelling of the covariance matrix, we need to convert the scale length and power spectrum exponent into a redshift-dependent galaxy-dark matter bias model according to Eq.~\eqref{eq:bias_conversion}. We use the aforementioned `global' correlation exponent model, as well as the assumption of linear biasing.
\paragraph*{Covariance modelling}
\label{para:covariance_modelling}
The modelling of our covariance matrix includes both the shot noise contribution and cosmic variance. However for simplicity we neglect correlations between angular modes and cross terms between the auto- and cross-correlations of the data vector. As shown in Fig.~\ref{fig:corr_matrix}, correlations between auto- and cross-correlations become important at small angular modes, i.e.\ larger scales. A more complete modelling should therefore include these terms.
\paragraph*{Cosmology dependence}
We must assume a fiducial cosmological model to convert from line-of-sight comoving distances to redshift. The dependence of the results on this assumption is expected to be mild \citep[see e.g.][]{0004-637X-684-1-88}. However in a more complete treatment, we will have to marginalize over cosmological parameters in the cross correlation analysis. Furthermore while the covariance matrix can, depending on the shot noise of the considered sample, show significant correlation between sub-degree and degree scales. This poses a problem if small scales are used to obtain cross correlation redshift posteriors that are used at larger scales in a cosmological analysis. We then need to consider a conditional likelihood of the large scale clustering measurement, given the small scale data vector. This is left for future work.
\paragraph*{Galaxy Sample Selection}
In this work we use a very simplistic selection of galaxy samples based on their stellar mass. Furthermore we assumed that we will be able to control the error contribution from uncertainties in the galaxy-dark matter bias of the spectroscopic sample to sufficient accuracy to not significantly widen our redshift posteriors. In future applications to data, we will have to include the spectroscopic measurement into the data likelihood and incorporate the complex galaxy selection function of spectroscopic surveys into the modelling. Especially at higher redshift, we must account for inhomogeneous galaxy populations in the spectroscopic reference sample and a more complex galaxy-dark matter bias model than presented in this work.
}

\section{Statistical Modelling}
\label{subsec:stats_modelling}
\begin{figure*}
\begin{center}
\includegraphics[width=\linewidth]{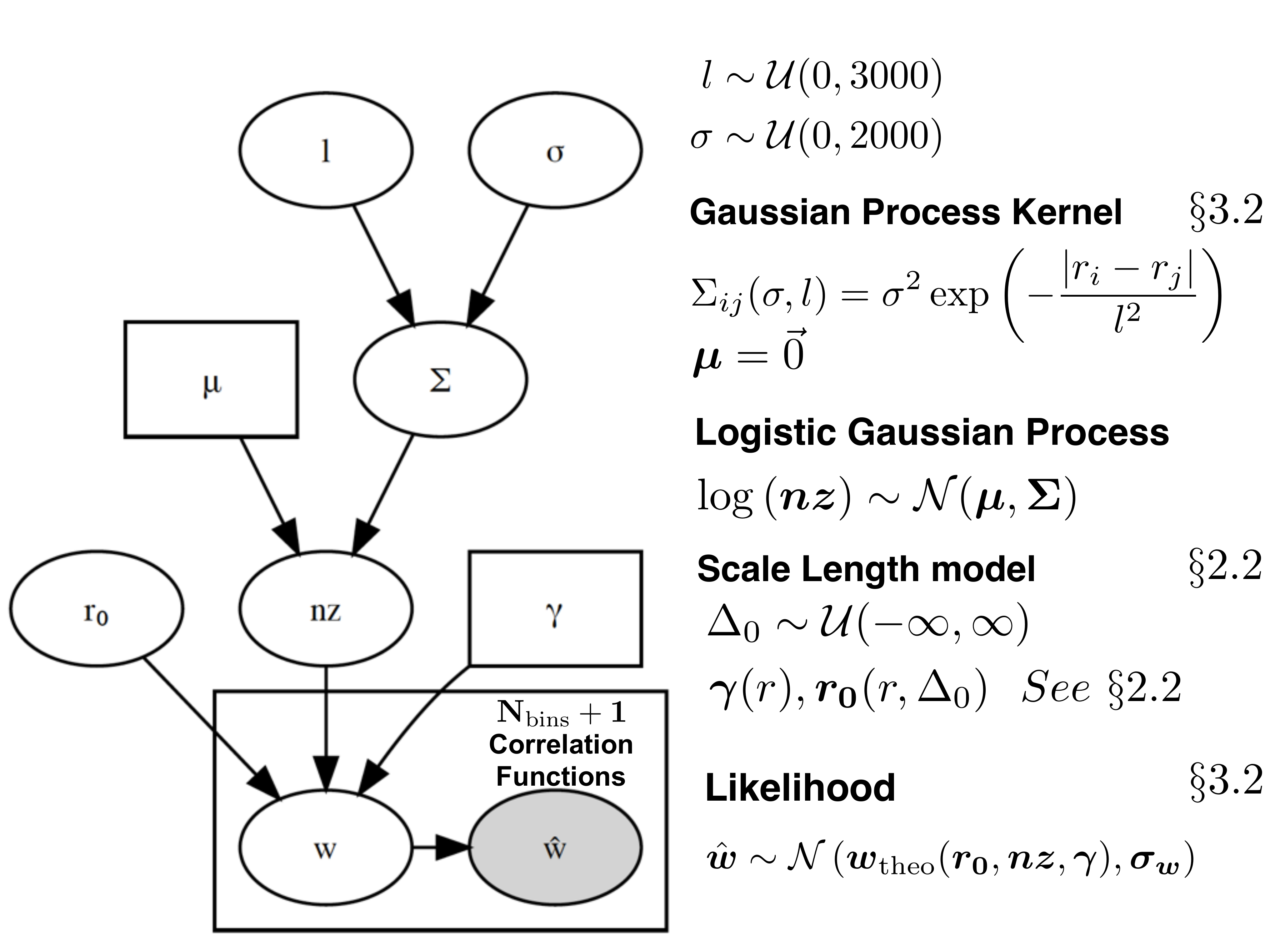}
\caption{\label{lab:diagram_wtheta} Graphical representation of the hierarchical logistic Gaussian process model. Empty/filled circles show unobserved/observed random variables and boxes denote fixed values. Arrows illustrate relationships between the variables. The boxed area represents a joint likelihood of angular correlation function measurements. This includes the autocorrelation of the photometric redshift distribution and the cross-correlations with the spectroscopic tophat bins. {\color{black} We list a summary of the model in the right column. Note that vector-valued quantities are shown in bold face. We also refer to the corresponding paper sections for more information. Here $\mathcal{U}(x, y)$ and $\mathcal{N}(\mu, \sigma)$ respectively denote a uniform distribution in the limits $[x, y]$ and a normal distribution with mean $\mu$ and standard deviation $\sigma$ or covariance $\Sigma$ in the multivariate case.   }}
\end{center}
\end{figure*}
In this section we will introduce our statistical model and discuss our sampling methodology. We use a graphical notation in the form of a directed graphical model shown in Fig.~\ref{lab:diagram_wtheta}. This diagram summarizes the dependency between the different random variables in a concise manner.
\subsection{Directed Graphical Models}
A directed graph, such as Fig.~\ref{lab:diagram_wtheta}, illustrates the dependency between different random variables schematically as a set of boxes or circles that are connected with arrows. Boxes denote variables with fixed values, circles represent random variables and shading represents a known random variable, e.g. a data vector. Arrows represent dependencies between these variables. For example, if two random variables $a$ and $b$ are connected by an arrow $a \rightarrow b$, they are not independent, i.e. $p(a,  b) \neq p(a) p(b)$ and we need to define the conditional probability $p(a | b)$ of a random variable $a$ given the value of $b$. The likelihood is shown in a box that represents the dimension of the random variable, in our case of the data vector {\color{black} of $N_{\rm bins} + 1$ angular correlation functions.}  A filled circle within such a box thus represents the measured data, i.e.\ the angular correlation function measurement $\widehat{w}$, where the dimension is specified in the larger box. The modelled data vector is again a random variable, as it depends on a set of parameters that are themselves random variables. Accordingly the model for the measured data, i.e.\ $w$, is shown as an open circle, where the model depends on a set of model parameters. The boxed nodes shown in Fig.~\ref{lab:diagram_wtheta} therefore represent the likelihood term $\mathcal{L} = p(\widehat{w} | w)$, i.e.\ the probability of the data given the model. In the following we will describe the different components of the particular model shown in Fig.~\ref{lab:diagram_wtheta}.

\subsection{Logistic Gaussian processes for Redshift Inference}
We model the redshift distribution of the photometric sample as a histogram in comoving distance shown in Fig.~\ref{lab:galaxy_dm_bias}, where the histogram bin midpoints are placed at the position of the $N_{\rm bins}$ spectroscopic tophat bins that are shown in red. We note that the bin sizes are selected such that they span equal-sized bins in redshift of size $\Delta z = 0.1$. We discretize the photometric redshift distribution, the distance dependent scale factor $r_0(r)$ and the exponent $\gamma(r)$ on this grid to model the angular correlation functions as described in \S~\ref{subsec:modelling_angcorr}. The data vector $\widehat{w}$ therefore consists of the auto-correlation of the photometric redshift bin and the cross-correlations between the photometric redshift bins and the spectroscopic tophat distributions. In this paper we will treat each bin height of the photometric sample, and the first order term in the Chebychev expansion (`scale length offset $\Delta_0$'), as random variables. We fix the scale length model of the spectroscopic sample to its fiducial value and marginalize over the scale length offset $\Delta_0$ of the photometric scale length model.

We note that the data covariance of the cross-correlation measurements implicitly contains our uncertainty in this parameter via the spectroscopic auto-correlation terms in Eq.~\eqref{eq:cov_matrix_ell}. The scale length models of the spectroscopic and photometric samples are strongly degenerate via the linear biasing assumption Eq.~\eqref{eq:linear_biasing}. For the considered data vector it is therefore practical to fix these spectroscopic terms, that will likely be constrained to good precision by future spectroscopic surveys like DESI. A more complete treatment would explicitly break this degeneracy by including the measurement of the spectroscopic correlation functions into the data vector. We leave this for future work.

\paragraph*{Likelihood specification}
We assume a Gaussian likelihood for the data vector $\widehat{w}$
\begin{equation}
    p(\widehat{w} | w_{\rm model}, \mathcal{C}) = \prod_{i=1}^{N_{\rm bins} + 1} \mathcal{N}(\widehat{w}_i | w_{\rm model, i}, \sigma_{\rm w, i}) \, ,
    \label{eq:cross_corr_like}
\end{equation}
where the product runs over the auto-correlation and the $N_{\rm bins}$ cross-correlations with the spectroscopic tophat bins.
The modelling of the angular correlation function is then given by Eq.~\eqref{eq:w_theta_binned}.

We illustrate the likelihood in Fig.~\ref{lab:diagram_wtheta} as the boxed node at the lower right corner, where the data/model of the angular correlation function is represented as the empty/filled nodes. The box represents the joint likelihood of the $N_{\rm bin} + 1$ dimensional data vector.
\paragraph*{Prior specification}
We set a logistic Gaussian process prior on the $N_{\rm bins}$ photometric redshift histogram heights denoted as the empty circle `nz' in Fig.~\ref{lab:diagram_wtheta}. Accordingly, the logarithm of the $N_{\rm bins}$ dimensional vector of histogram heights $\log{\left(\mathbf{nz}\right)}$ is assumed to be drawn from a multivariate normal distribution with mean vector $\mathbf{\mu}$ and $N_{\rm bins} \times N_{\rm bins}$ dimensional covariance matrix $\Sigma$
\begin{equation}
    \log{\left(\boldsymbol{nz}\right)} \sim \mathcal{N}\left(\boldsymbol{\mu}, \boldsymbol{\Sigma}\right) \, .
\end{equation}
{\color{black} The advantage of using a logistic Gaussian Process prior -- in contrast to alternatives such as the Dirichlet or flat priors -- is that it opens the possibility of encoding knowledge of the shape and smoothness of the redshift distribution in the covariance function. This is however optional, and the Gaussian can be chosen to be very uninformative.

As described in the following, we can jointly marginalize over the parameters that describe the redshift distribution and the scale length model, which accounts for the degeneracy between the parameters of the redshift distribution and the galaxy-dark matter bias model. Furthermore, the logistic transform guarantees positivity in the $\mathbf{nz}$ posteriors. These aspects are an advantage over applying a Gaussian Process interpolation, e.g., on the posteriors from a clustering redshift technique.    }

If not mentioned otherwise, we fix the mean `$\boldsymbol{\mu}$' of the logistic Gaussian process to zero and marginalize over the uncertainty in the covariance matrix. For this we choose the following kernel
\begin{equation}
    \mathbf{\Sigma}_{i, j}(\sigma, l) = \sigma^2 \exp{\left(- \frac{|r_i - r_j|}{l^2}\right)} \, ,
\end{equation}
{\color{black} which is a special case of the Mat{\'e}rn kernel \citep{Rasmussen2006Gaussian}, a common choice for Gaussian processes.}
We also tried the squared exponential kernel, however we found that this choice produced numerically more stable covariance matrices.
The two parameters $\sigma$ and $l$ govern the magnitude of the diagonal components, as well as the correlation, or smoothness, of the histogram. In our model we therefore add a hierarchy that marginalizes over these parameters using broad uniform priors $l \sim \mathcal{U}(0, 3000)$ and $\sigma \sim \mathcal{U}(0, 2000)$. {\color{black} The units of these parameters are $[l] = \sqrt{{\rm Mpc}}$ and $[\sigma] = {\rm Mpc}$ respectively.} These random variables, denoted `l' and `$\sigma$' are shown in Fig.~\ref{lab:diagram_wtheta} in the top hierarchy.
As detailed in \S \ref{subsec:modelling_angcorr}, we fix the redshift dependence of the power law exponent `$\gamma$' and choose an unrestricted flat prior on the first coefficient of the Chebychev expansion, the `scale length offset $\Delta_0$'.
\subsection{Sampling Methodology}
\label{sec:analysis_results}
We sample the model shown in Fig.~\ref{lab:diagram_wtheta} using a combination of Metropolis-Hastings sampling and Elliptical Slice sampling steps. {\color{black} We use two separate one-dimensional Metropolis-Hastings steps to sample the two parameters that describe the Gaussian Process Kernel, a one-dimensional Metropolis-Hastings step to sample the parameter that describes the scale length model and an elliptical-slice sampling step to sample the parameters that describe the background sample redshift distribution. Since in each of these steps we fix the other parameters fixed, which mimics Gibbs sampling, we will refer to the procedure as `Metropolis-Hastings within Gibbs' sampling.}  {\color{black} The Metropolis-Hastings sampling approach is described in \S~\ref{subsubsec:gibbs_sampling} and the Elliptical Slice sampling method in \S~\ref{subsec:elliptical_slice_sampling}.}  We will then detail the concrete sampling implementation in \S~\ref{subsubsec:sampling_log_gp}.

\subsubsection{Metropolis-Hastings sampling}
\label{subsubsec:gibbs_sampling}
Assuming a set of random variables $\mathbf{\theta}$, {\color{black} Metropolis-Hastings within Gibbs sampling \citep[see e.g.][]{gelmanbda04} iteratively samples from the distribution of each variable $\theta_{i}$ conditional on all other variables $\theta_{-i}$, i.e. $ p(\theta_i | \theta_{-i}^{t - 1}, \mathcal{D})$, where $\theta_{-i}^{t - 1} = \left(\theta_1^{t}, \dots, \theta_{i - 1}^t, \theta_{i+1}^{t - 1}, \dots, \theta_d^{t - 1}\right)$.} Here $\mathcal{D}$ denotes the data vector, $t$ counts the number of iterations and $i$ denotes the index of the variable that is currently updated. If this conditional distribution is not known, we draw samples for a new parameter $\theta^{\star}$ given a previous state of the parameter $\theta^{t-1}$ from a proposal distribution $J_t(\theta^{\star} | \theta^{t-1})$. These samples are then accepted with probability
{\color{black}
\begin{equation}
        r = \frac{p(\theta_{i}^{\star} | \theta_{-i}, \mathcal{D})/J_t(\theta_{i}^{\star} | \theta_{i}^{t-1})}{p(\theta_{i}^{t - 1} | \theta_{-i}, \mathcal{D})/J_t(\theta_{i}^{t-1} | \theta_{i}^{\star})} \, .
        \label{eq:metropolis_hastings}
\end{equation}}

We can connect the posterior probabilities $p(\theta_{i} | \theta_{-i}, \mathcal{D})$ via Bayes rule with the likelihood $p(\mathcal{D} | \theta_{i}, \theta_{-i})$ and the prior $p(\theta_i)$ as
\begin{equation}
   p(\theta_{i} | \theta_{-i}, \mathcal{D}) \propto p(\mathcal{D} | \theta_{i}, \theta_{-i}) \, p(\theta_i | \theta_{-i}) \, .
\end{equation}
We select a normal distribution as the proposal distribution $J_t(\theta_{i}^{\star} | \theta_{i}^{t-1}) = \mathcal{N}(\theta_{i}^{\star} | \mu=\theta_{i}^{t - 1}, \sigma)$, where we center the Gaussian at the previously proposed value $\theta_{i}^{t - 1}$ and tune the standard deviation $\sigma$ such that we obtain acceptance rates between 20\%-30\% for the respective parameters. Since this distribution is symmetric, it cancels in Eq.~\eqref{eq:metropolis_hastings}.

{\color{black} We note that samples are always accepted ($r = 1$) if we use the conditionals $ p(\theta_i | \theta_{-i}^{t - 1}, \mathcal{D})$ in Eq.~\ref{eq:metropolis_hastings} as the proposal distribution. This special case is known as Gibbs sampling and can be used for a restricted set of models, typically complex conjugate models, for which we can obtain an analytical distribution. We note that the Metropolis-Hastings steps described do not have to be one-dimensional (single parameter updates), but rather can use multidimensional proposal distributions.

Furthermore we note that the sampling algorithm used to update each parameter does not have to employ the Metropolis-Hastings scheme, but can use other sampling approaches. The iterative update of the conditionals then gives the possibility to use different sampling algorithms as `building-blocks' for a combined sampling scheme. We will describe an alternative scheme in the next section and} refer the interested reader to \citet{gelmanbda04} for a more detailed description of the Metropolis-Hastings sampling methodology.

\subsubsection{Elliptical Slice Sampling}
\label{subsec:elliptical_slice_sampling}
{\color{black} Slice sampling samples from a density $p(z)$ by considering the area under the density function represented by the joint density of $z$ and an auxiliary variable $u$. After sampling from this joint density, samples of the variable $u$ are dropped to obtain samples from $p(z)$. Following e.g. \citet{Dittmar_slicesampling} we can write:
\begin{equation}
    p(z) = \int_{0}^{\hat{p}(z)} \frac{1}{Z_{\hat{p}}} du = \int p(z, u) du \, ,
\end{equation}
where $\hat{p}(z)$ does not have to be normalized and $Z_{\hat{p}}$ is a normalization constant $\frac{\hat{p}(z)}{Z_{\hat{p}}} = p(z)$. The joint distribution is then given as
\begin{equation}
    p(z, u) = \begin{cases}
     1/Z_{\hat{p}} \ &\text{if} \ 0 \leq u \leq \hat{p}(z) \\
     0 \ &\text{otherwise}
   \end{cases} \, .
\end{equation}

Using the same methodology introduced in the previous section, we first sample from $p(u|z)$ using $u \sim \mathcal{U}\left[0, \hat{p}(z)\right]$ and then from $p(z | u)$ represented as a sample from the `slice' $\{z: u < \hat{p}(z)\}$.

While the first step of slice sampling is trivial, the second often involves starting with an initial sample and selecting a larger proposal region, determined by a step size parameter, that contains both the sample and the slice. Then an initial sample is drawn from that region. If this initial sample lies outside the target slice, the size of the proposal region is reduced and the procedure is repeated. Concretely we note that it is still necessary to select a step size parameter, to define the proposal region.

Elliptical Slice sampling is a variant especially geared towards target distributions of the form
\begin{equation}
    p^{\star}(\theta) = \frac{1}{Z} \, p(\mathcal{D} | \theta) \, \mathcal{N}(\theta, 0, \Sigma)  \, ,
\end{equation}
where $\theta$ denotes the free parameters of the model, $ \mathcal{N}(\theta, 0, \Sigma)$ is a Gaussian prior and $p(\mathcal{D} | \theta)$ describes the likelihood. In contrast to classical slice sampling, the proposal region is now an ellipse defined by a sample from the prior $\nu \sim \mathcal{N}(\nu, 0, \Sigma)$ and the current parameter state $\theta^{t - 1}$:
\begin{equation}
    \theta^{*} = \theta^{t-1} \cos\left(\alpha\right) + \nu \sin\left(\alpha\right) \, .
\end{equation}
Here, $\alpha$ describes the position on this elliptical proposal region.  Since the initial sample is drawn from the full ellipse, it is not necessary to select a step size. The sampling and shrinking operations are then performed in analogy to the classical slice sampling algorithm.
}
We refer the interested reader to \citet{2010arXiv1001.0175M} for a more detailed description of the method.

\subsubsection{Sampling the Logistic Gaussian process}
\label{subsubsec:sampling_log_gp}
Starting from the top end of Fig.~\ref{lab:diagram_wtheta}, we use three main Metropolis-Hastings steps to sample the hyperparameters of the logistic Gaussian process covariance, the (log) histogram heights of the photometric redshift distribution and the scale length offset $\Delta_0$ that describes the scale length model. The sampling of the hyperparameters $l$ and $\sigma$ is performed in two separate Metropolis-Hastings steps. Then, we sample the histogram heights using Elliptical Slice sampling. In the third Metropolis-Hastings step we sample the scale length offset $\Delta_0$. In each of these steps we hold the parameters that are not currently updated fixed, as described in \S~\ref{subsubsec:gibbs_sampling}.

We note that the parameters that describe the histogram heights and scale length model can be quite correlated and Elliptical Slice sampling allows us to take larger steps along the degeneracy direction, which leads to better sampling performance for the logistic Gaussian process model. We further note that this methodology does not require gradients, which will not be available if we sample over a likelihood, which models the correlation functions in terms of cosmological parameters.

In the following we will discuss how we construct a prior on the redshift distribution from a Bayesian histogram of sampled redshift values. In the experiments we performed in the next section we found that Metropolis-Hastings sampling performed better in combination with this redshift prior. The reason is that this prior weakens the correlation between the aforementioned parameters, in which case Metropolis-Hastings sampling becomes more efficient. We therefore used Metropolis-Hastings sampling steps for the individual log histogram heights instead of Elliptical Slice sampling. We note that this might not be an optimal choice for different priors and data likelihoods.
\subsubsection{Informative Prior on the Redshift distribution}
\label{subsec:prior}
To set an {\color{black} informative} prior on the redshift distribution, we require an independent set of data that constrains the histogram bins represented as the logistic Gaussian process. This information will typically come from photometric redshift estimation based on template fitting or machine learning, that constrain the redshifts of individual galaxies. A prior on the population of redshift values can therefore be represented as a Bayesian histogram and samples can be drawn from a Dirichlet distribution given the counts of galaxy redshifts in the respective histogram bins. The distribution over the set of histogram heights $\widehat{nz}$ is therefore given as
\begin{equation}
    \widehat{nz} \sim {\rm Dir}(\Phi + \widehat{c}) \, ,
\end{equation}
where {\rm Dir} denotes the Dirichlet distribution and $\Phi$ its parameter that is set to a unity vector to represent a flat prior. The vector $\widehat{c}$ denotes the counts of galaxy redshifts in the respective histogram bin. Here $\widehat{c}$ is a vector of length $M$, where $M$ denotes the number of histogram bins. Accordingly, a sample  $\widehat{nz}$ from the Dirichlet distribution is an $M$ dimensional vector of histogram heights. If the sample size is large, i.e.\ if {\color{black} each element of} $\widehat{c}$ is high, the scatter of new samples around the mean histogram heights is small. However if there are very few galaxies in the sample, the Dirichlet distribution will sample around this mean value of histogram heights more broadly.
Since the treatment of photometric redshift error in the context of machine learning or template based photometric redshift techniques is beyond the scope of this paper, we will use a lower bound on the prior width by assuming true redshift values for all galaxies in the photometric dataset.

For our case, we find that the logarithm of these sampled redshift histogram heights can be approximated as a multivariate normal distribution to sufficient accuracy for our purposes. For this we use the sample mean and covariance estimated on a large number ($>10^8$) of (log) Dirichlet samples for our approximation. This multivariate normal distribution can then be readily used as a prior distribution for the logistic Gaussian process model.

\subsection{Model Evaluation}
In many practical application of inference there will be a level of uncertainty in the underlying modelling and, as a result, a level of ambiguity between reasonable model choices. It is therefore of paramount importance to compare these models efficiently and combine them if various models provide equally good descriptions of the data.
We evaluate and compare the models using goodness of fit statistics and the expected biases in the Lensing convergence power spectrum. The latter is especially useful as it allows judgment about the possible impact of residual photometric redshift errors on the weak lensing measurement, which is the primary cosmological probe to test theories of dark energy in the context of large area photometric surveys.

\subsubsection{Information Criteria}
In classical statistics the maximum of the log-likelihood is often penalized by the number of free parameters in the model, to measure the `goodness of fit' and avoid the risk of specifying too many parameters (`overfitting'). This approach, represented by the Akaike information criterion  \citep[AIC;][]{akaike1973information}, has the disadvantage in the Bayesian analysis, that it will not change with different prior choices that clearly affect model complexity and posterior. In this work we will therefore use the Deviance Information Criterion \citep[DIC;][]{spiegelhalter2002bayesian} that provides an extension of the AIC towards the Bayesian paradigm.

We use the definition of the DIC based on the predictive density\footnote{{\color{black} The original definition based on the deviance differs slightly by a factor of -2. This is further explained in \citet{2013arXiv1307.5928G}, where both definitions are detailed.}} \citep{2013arXiv1307.5928G} as
\begin{equation}
    {\rm DIC} = \log{\left(p\left(\mathcal{D} | \widehat{\theta}_{\rm Bayes}\right)\right)} - p_{\rm DIC}  \,
    \label{eq:dic_def}
\end{equation}
where $\mathcal{D}$ denotes the data vector and  $p\left(\mathcal{D} | \widehat{\theta}_{\rm Bayes}\right)$ denotes the likelihood evaluated at the posterior expectation
\begin{equation}
    \widehat{\theta}_{\rm Bayes} = E\left[\theta | \mathcal{D}\right] \, .
\end{equation}
Model complexity is then penalized as
\begin{equation}
     p_{\rm DIC} = 2 \, {\rm var}_{\rm Post}{\log{\left(\mathcal{D} | \theta\right)}} \, ,
     \label{eq:pdic}
\end{equation}
where ${\rm var}_{\rm Post}{\log{\left(\mathcal{D} | \theta\right)}}$ denotes the posterior variance of the likelihood. More complex models will have a larger posterior variance due to their additional degrees of freedom in parameter space. As both terms in the DIC are evaluated with respect to the posterior, its value will depend on the choices of prior.

{\color{black} We note that an alternative to this approach (that is commonly used in cosmology) is a model selection using the marginal likelihood, or evidence, for a model $\mathcal{M}_i$ from a set of $C$ models $\mathcal{M} = \{\mathcal{M}_1, \dots, \mathcal{M}_{C}\}$:
\begin{equation}
 p(\mathcal{D} | \mathcal{M}_i) = \int p(\mathcal{D} | \mathbf{\theta}, \mathcal{M}_i) \, p(\mathbf{\theta} | \mathcal{M}_i) \, d\theta \, ,
 \label{eq:bma}
\end{equation}
where $\theta$ are the model parameters and $\mathcal{D}$ represents the data. While we could have used multimodel inference techniques like \citet{doi:10.1080/00031305.2013.791644} on our previously fitted chains to estimate marginal likelihood for all the models used in this paper, we decided to use the ${\rm DIC}$ criterion for its simplicity and computational speed.

We note, however, that a Bayesian information criteria like the DIC can have advantages when averaging over multiple models becomes necessary, or if multiple models need to be `scored' based on their `goodness of fit'. The evidence can be used to `weight' the posterior constraints for different models in the case where the true model is known to be in the model set $\mathcal{M}$. If this is not the case, this approach will select the single `best fit' model in the large data limit. In cases where the true model is not part of the model set, techniques based on information criteria or `stacking' techniques \citep[e.g.][]{2013arXiv1307.5928G} may be more suitable. Furthermore the evidence can be dependent on prior choices \citep{2013arXiv1307.5928G}, which may lead to specification problems especially if the model set is incomplete. Thus, while our primary motivation for using the computationally more efficient DIC criterion is computational speed and simplicity, we would like to highlight the usefulness of information criteria, especially if we expect systematic modelling errors that could cause an incomplete model set.
}
\subsubsection{Bias in lensing convergence power spectrum}
To be able to better evaluate the significance of photometric redshift biases, we evaluate the quality of the reconstructed photometric redshift distribution in terms of the expected relative bias in the lensing convergence power spectrum
\begin{equation}
    \Delta_{\rm rel} =  \frac{\widehat{C}_{\ell}^{\rm fid.} - \widehat{C}_{\ell}^{\rm bias}}{\widehat{C}_{\ell}^{\rm fid.}} \, ,
    \label{eq:lensing_bias_def}
\end{equation}
where $\widehat{C}$ denotes the binned lensing power spectrum
\begin{equation}
    \widehat{C}_{\ell} = \frac{1}{N_{\ell}} \sum_{\ell} C_{\ell} \, ,
\end{equation}
and $N_{\ell}$ denotes the number of integer angular modes. Here, $\widehat{C}_{\ell}^{\rm fid.}$ denotes the lensing convergence power spectrum obtained using the unbiased, fiducial, redshift distribution. $\widehat{C}_{\ell}^{\rm bias}$ denotes the photometric redshift distribution obtained assuming the (potentially) biased galaxy-dark matter bias model in the inference. We choose to bin the angular correlation power spectra $\ell \in [10, \ell_{\rm max}]$, where we select $\ell_{\rm max} = 3000$ in analogy to \citet{2015MNRAS.451.4424K}.
We compare the resulting systematic biases with the expected statistical measurement error from a weak gravitational lensing measurement with the specifications given in Tab.~\ref{tab:cosmo_param} and a shape noise of $\sigma_{\epsilon} = 0.23$ comparable with \citet{2015MNRAS.451.4424K}. The error in the binned measurement is then
\begin{equation}
    {\sigma}(\widehat{C}_{\ell}) = \frac{1}{N_{\ell}} \sqrt{\sum_{\ell} \frac{2}{(2 \ell + 1) f_{\rm sky}}\left(C_{\rm \ell} + \frac{\sigma_{\epsilon}^2}{2 \overline{n}_g}\right)^2} \, ,
    \label{eq:stat_err_lens}
\end{equation}
which implements Eq.~\eqref{eq:cov_matrix_ell} with a different, weak gravitational lensing specific, shot noise term. {\color{black} Eq.~\eqref{eq:stat_err_lens} also includes the cosmic variance contribution to the covariance matrix, that is inversely proportional to the fractional sky coverage $f_{\rm sky}$. Neglected are the covariance between the different modes $\ell$. This is done in analogy to other forecasts like \citet{2015MNRAS.451.4424K}.} The $1 \ \sigma$ error in $\Delta_{\rm rel}$ is then given as
\begin{equation}
    \sigma_{\Delta_{\rm rel}} = \frac{{\sigma}(\widehat{C}_{\ell})}{\widehat{C_{\ell}^{\rm fid}}}
    \label{eq:err_delta}
\end{equation}

We note that this simple metric does not substitute an accurate forecast that evaluates the impact of photometric redshift bias on cosmological parameter inference. This would require a tomographic analysis that is representative of modern surveys like DES or LSST, which is beyond the scope of this work. However, if the mean systematic bias in the reconstructed {\color{black} lensing convergence power spectrum} is significantly below the statistical error budget given in Eq.~\eqref{eq:stat_err_lens}, we can assume that its impact on cosmological parameters will be small. It therefore allows us to approximately  judge the practical implications and relevance of the photometric redshift error on cosmological parameter inference, even without a more precise forecasting exercise.
\section{Analysis and Results}
\label{sec:results}
In the following section we investigate the performance of our logistic Gaussian process model in recovering the photometric redshift distribution using the cross-correlation likelihood (Eq. \ref{eq:cross_corr_like}) and test how different scale-length models affect the inference.

Tab.~\ref{tab:bias_models} summarizes the nomenclature used in our analysis. The first column quotes the name tag of the particular setup, the second column the scale length model assumed for the spectroscopic sample, the third our ansatz for the scale length model of the photometric sample and the fourth column lists the true scale length model of the photometric sample (that is unknown to us in practise). The final column indicates if a prior on the photometric redshift distribution is used. We refer to Fig.~\ref{lab:galaxy_dm_bias} for an overview over the different scale length models. We note that we adapt the covariance matrix of the likelihood to changes in the corresponding galaxy-dark matter bias according to Eq.~\eqref{eq:bias_conversion}.

\subsection{Fiducial Model}
\label{subsec:fid_model}
In this first section we study a test case where our assumption that the scale length model of the photometric and spectroscopic samples only differs by a constant offset is valid to a good degree. This is the case for example in Model 4 and 5 shown in Fig.~\ref{lab:galaxy_dm_bias}. As can be seen this implies that both samples have similar stellar mass ranges, which would require a spectroscopic survey that does not exhibit an extreme selection towards highly biased tracers. We therefore study this scenario assuming Model 4 as the scale length model of the photometric sample and Model 5 as the scale length model of the spectroscopic sample. As discussed in \S~\ref{subsec:modelling_galax_bias}, we use a constant offset to parametrize our uncertainty in the scale length, which corresponds to the first (constant) term in the Chebychev expansion. In the following, we will refer to this case as `S5A5P4'.

To test if the small differences between Model 4 and 5 will introduce biases in our inference, we compare with the results in the absence of systematics. For this unbiased, fiducial test case, we fix the higher order terms of the Chebychev expansion to their true, unbiased, values from Model 4. As before, we will marginalize over an offset in the scale length and will refer to this case as `S5A4P4'. Note that this scenario is not the same as assuming scale length model 4 for both the spectroscopic and the photometric sample, because our covariance matrix will assume scale length model 5 for the spectroscopic sample. To simplify the notation, the following will refer to the offset in the scale length as `scale length offset $\Delta_0$'.

\begin{figure*}
\begin{center}
\includegraphics[width=\linewidth]{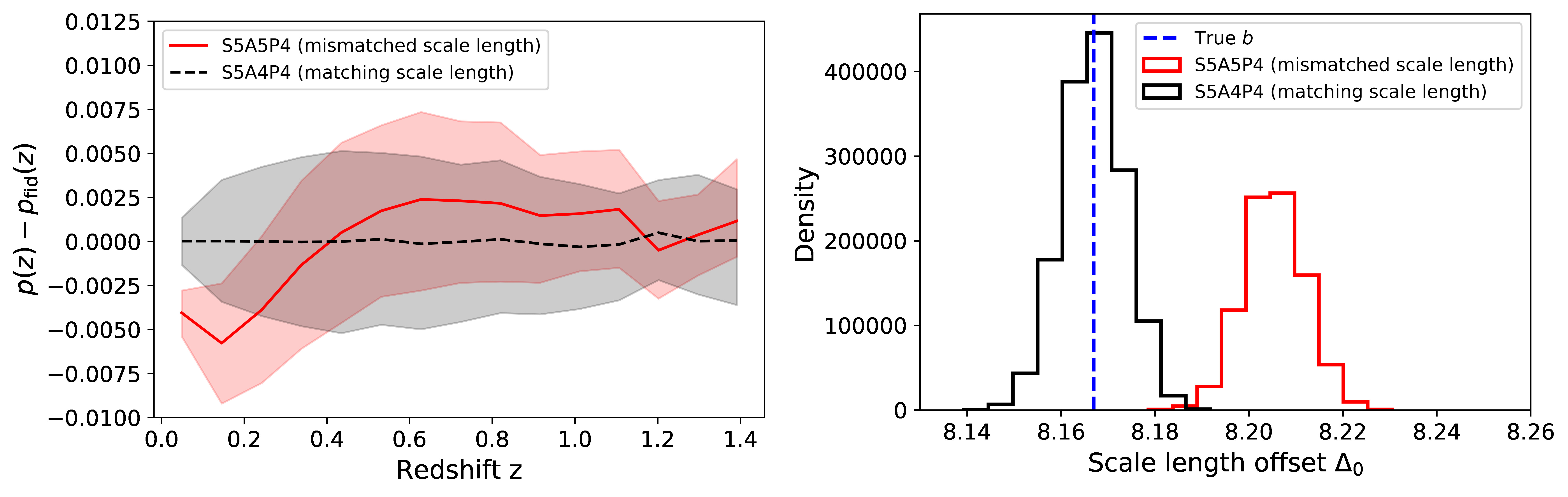}
\caption{\label{lab:fid_results} {\color{black} Posterior distributions for the bin heights of the ensemble redshift distribution of the photometric sample and scale length offset $\Delta_0$ for the model with mismatched scale length S5A5P4 (red) and matching scale length S5A4P4 (black/grey). \textit{Left:} We show the $[5, 95]$ posterior percentiles of the difference between the posterior redshift distribution and the true redshift distribution. \textit{Right:} Posterior distributions of the `Scale length offset $\Delta_0$'. The blue dashed line shows its true value.}}
\end{center}
\end{figure*}

The left panel of Fig.~\ref{lab:fid_results} shows the posterior of the difference between the true photometric redshift distribution and the redshift distribution inferred in these models. All distributions shown in this work have been normalized to unit area. We show the results for `S5A4P4' in grey contours and results for `S5A5P4' in red. The right panel plots the corresponding posterior distributions for the scale length offset $\Delta_0$, where the dashed blue line denotes its true value.
We see that we can recover the redshift distribution to great accuracy for both models. For the unbiased scale length model we obtain, as expected, unbiased constraints on the photometric redshift distribution and scale length parameter, which we regard as a test of our sampling and methodology.

The $[5, 95]$ posterior percentile range of the inferred photometric redshift distribution assuming the slightly biased case S5A5P4 are largely consistent with the true result, however, as expected, inconsistent for the scale length offset $\Delta_0$. We can therefore conclude that a slight modelling bias in the scale length, or galaxy-dm bias model, has a small effect on the accuracy of the recovered redshift distribution. It has to be noted that we make quite optimistic assumptions for the photometric and spectroscopic surveys (see Tab.~\ref{tab:cosmo_param}). For less optimistic assumptions, e.g. a smaller area of overlap between spectroscopic and photometric survey, we can expect the statistical error to be larger. The statistical error then quickly becomes even more dominant compared with the small photometric redshift systematic shown in Fig.~\ref{lab:fid_results}.

\begin{figure}
\begin{center}
\includegraphics[width=\linewidth]{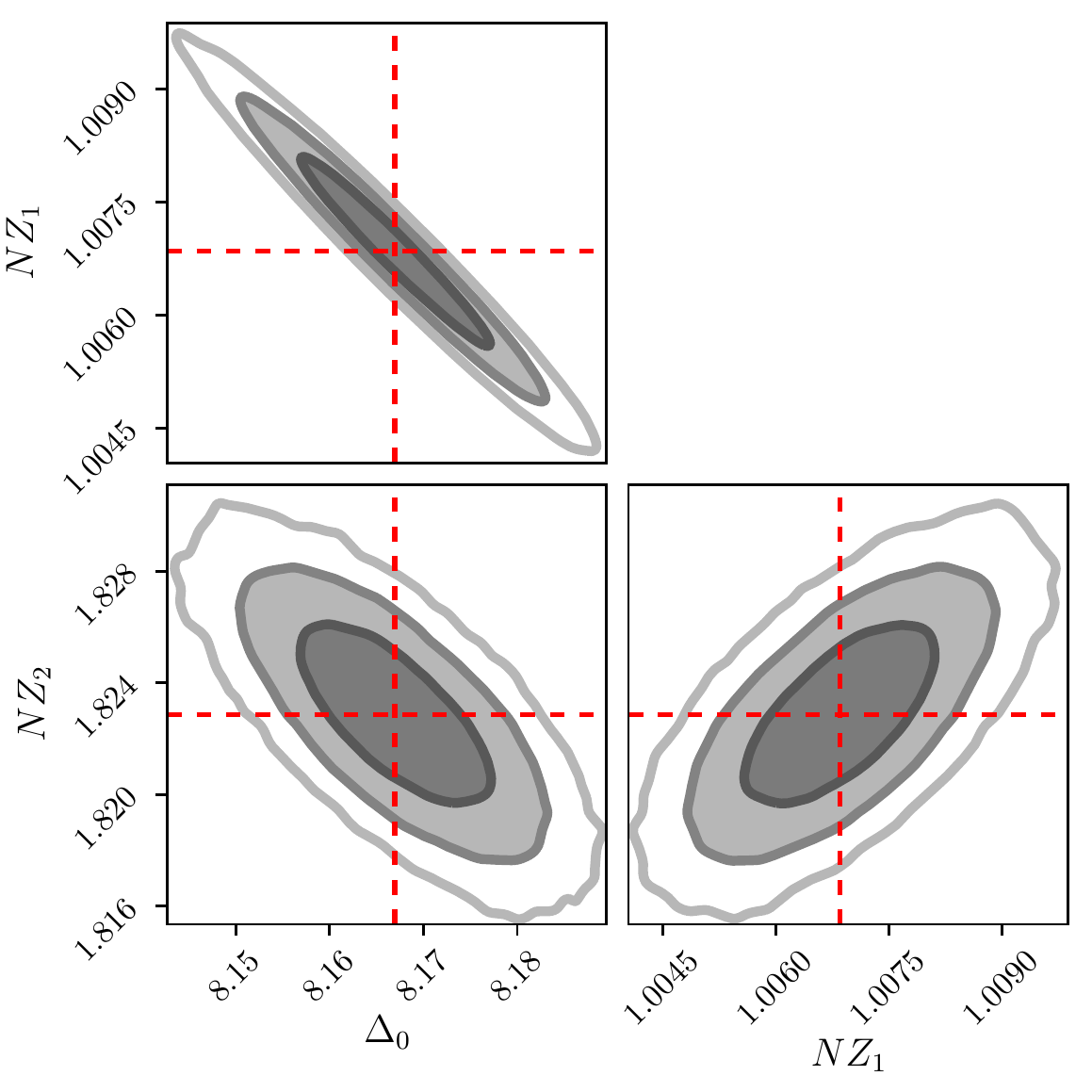}
\caption{\label{lab:post_constant_bias} Posterior distributions for the unbiased model {\color{black} (matching scale length)} S5A4P4 for the scale length offset $\Delta_0$ and the first two photometric redshift bins. The dashed lines denote the true values.  }
\end{center}
\end{figure}

We note that there is considerable degeneracy between the scale length offset and the redshift distribution, particularly for the low redshift bins where the signal-to-noise ratio in the clustering measurement is large. This is illustrated in Fig.~\ref{lab:post_constant_bias} for the unbiased model S5A4P4, which shows the posterior distributions of the scale length offset $\Delta_0$ and the first two histogram bin heights of the photometric redshift distribution. We see that the scale length offset $\Delta_0$ and the photometric redshift distribution bin heights are indeed strongly correlated. For the bins at higher comoving distance, or redshift, this correlation decreases.

In the next section we will discuss a more extreme case, where larger differences in the stellar mass of the photometric and spectroscopic galaxy population translate into larger differences in the scale length models.
\begin{table}
\centering
\caption{ Summary of the model nomenclature used in this work.  We list the model name, the assumed scale length model (see Fig.~\ref{lab:galaxy_dm_bias}) for the spectroscopic sample, the Ansatz for the scale length of the photometric sample and the true scale length model of the photometric sample. The last column indicates if a prior on the photometric redshift distribution was used. }
\label{tab:bias_models}
\begin{tabular}{l  l  l  l  l   }
Name                 & Spec & Ansatz & Phot & Prior \\\hline\hline
S5A4P4       & 5    &    4   &  4   & No \\\hline
S5A5P4                & 5    &    5   &  4   & No \\\hline
S1A2P4         & 1    &    2   &  4   & No \\\hline
S1A5P4       & 1    &    5   &  4   & No  \\\hline
S1A1P4               & 1    &    1   &  4   & No \\\hline
S1A2P4 prior Nz & 1    &    2   &  4   & Yes  \\\hline
S1A5P4 prior Nz & 1    &    5   &  4   & Yes
\end{tabular}
\end{table}
\begin{table}
\centering
\caption{Results of the model comparison between different redshift dependent galaxy-dark matter bias models summarized in Tab. \ref{tab:bias_models}. DIC denotes the Deviance Information criterion (Eq. \ref{eq:dic_def}), $p_{\rm DIC}$ the measure of model complexity (Eq. \ref{eq:pdic}), `Loglike' denotes the log-likelihood evaluated at the mean of the posterior, i.e. the first term in Eq.~\eqref{eq:dic_def}. `Cl Bias' denotes the relative bias in the lensing convergence power spectrum (Eq. \ref{eq:lensing_bias_def}). }
\label{tab:model_comparison}
\begin{tabular}{l  l  l  l  l   }
Name            & DIC & Cl Bias & Loglike & $p_{\rm DIC}$ \\\hline\hline
S1A2P4 & 77.85    & 1.77 & 95.71 & 17.86 \\\hline
S1A5P4 & 81.54   & 0.26 & 95.58 & 14.04 \\\hline
S1A1P4 & -2385.84 & 3.70 & -2370.28 & 15.57 \\\hline
S1A2P4 prior Nz & -117.47 & 0.76 & 24.31 & 141.78 \\\hline
S1A5P4 prior Nz & 93.30 & 0.10 & 101.05 & 7.75
\end{tabular}
\end{table}
\begin{figure}
\begin{center}
\includegraphics[width=\linewidth]{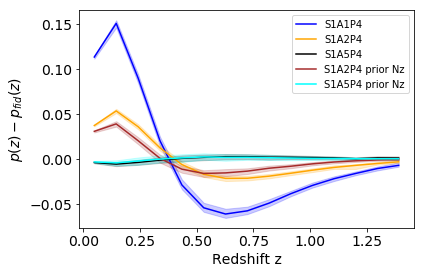}
\caption{\label{lab:comparison_nz} Difference between the estimated $p(z)$ and fiducial $p_{\rm fid}(z)$ redshift distributions for the different scenarios listed in Tab.~\ref{tab:bias_models}. The solid lines show the median of the posterior and the error contours the $[5, 95]$ posterior percentiles.}
\end{center}
\end{figure}
\begin{figure}
\begin{center}
\includegraphics[width=\linewidth]{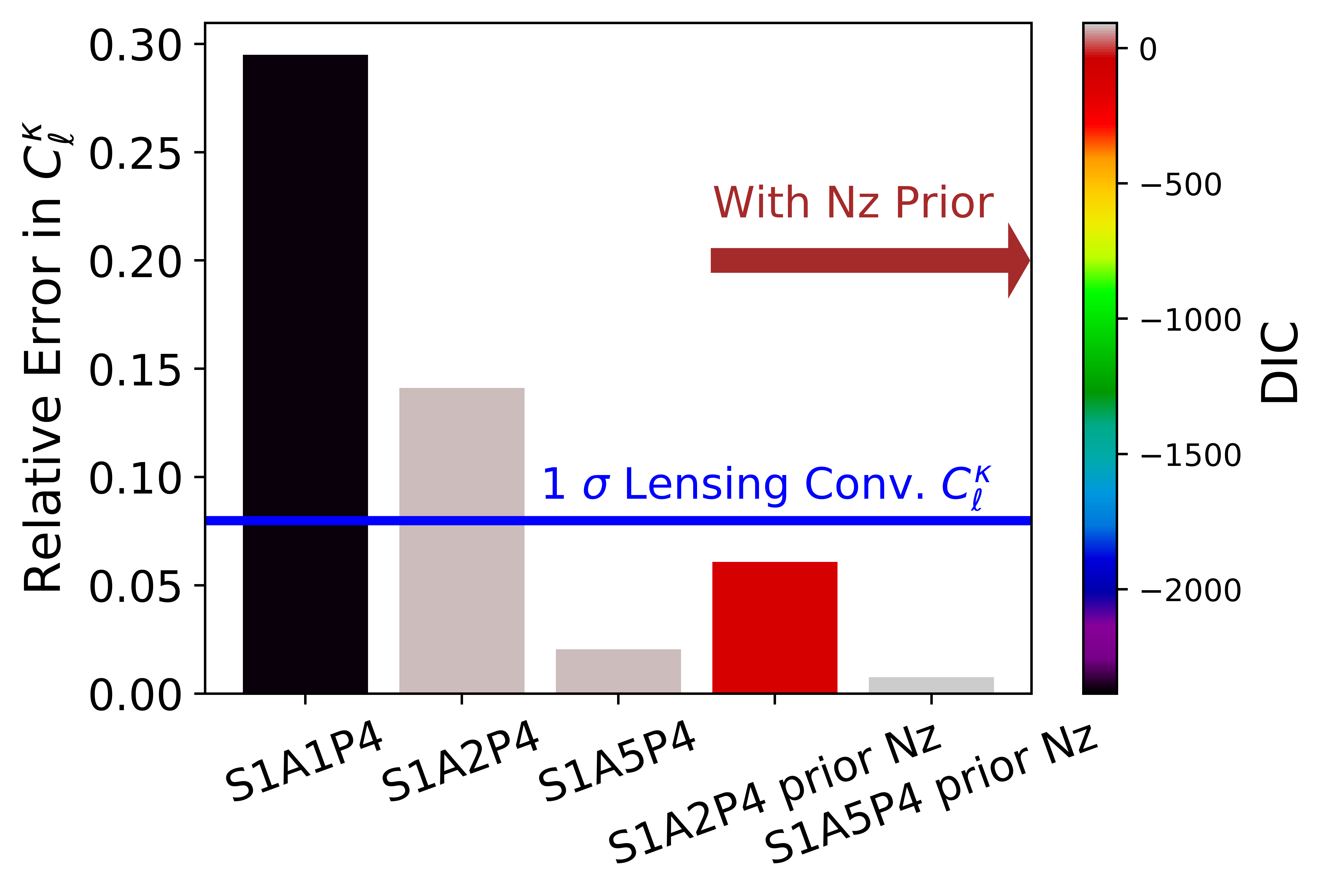}
\caption{\label{lab:model_comparison} Summary of model complexity, as measured by the DIC (Eq.~\ref{eq:dic_def}), and relative error in the lensing convergence power spectrum (Eq.~\ref{eq:lensing_bias_def}) for different scale length model combinations (see Tab.~\ref{tab:bias_models}). The bar heights show the relative error in the lensing convergence power spectrum for the different scenarios, the color their DIC (Eq.~\ref{eq:dic_def}) values and the blue horizontal line the $1 \ \sigma$ statistical error budget in the lensing correlation power spectrum given by Eq.~\eqref{eq:err_delta}.  }
\end{center}
\end{figure}

\subsection{Impact of biased scale length models}
\label{subsec:impact_wrong_scale_length}
We test the impact of biased scale length models on the redshift distribution inference in a more pessimistic setup: we {\color{black} make the ansatz} that the scale length of the photometric/spectroscopic sample is given by scale length Model 4/1 and denote this setup as S1A1P4. As one can see from Fig.~\ref{lab:galaxy_dm_bias}, a simple constant shift in the scale length model is no longer a sound approximation, as both models have a significantly different slope at large comoving distances or high redshift. This can be seen as an extreme case of scale length model mis-specification and we will likely be able to calibrate the model to better accuracy either using simulations or by incorporating information from e.g. galaxy-galaxy lensing \citep[e.g.][]{2018PhRvD..98d2005P} in future surveys.

We therefore include a more optimistic setup into this comparison. Here we make the ansatz for Models 2/5, that are more similar to the true photometric model (Model 4). This means that instead of fixing the higher order coefficients of the Chebychev expansion to the parameters of Model 1, we fix them to the parameters of Model 2/5. We will denote the Model 1-4 with an ansatz of Model 2/5 as `S1A2P4' and `S1A5P4' respectively. As in the previous section we will marginalize over the constant term in the Chebychev expansion, i.e. the `scale length offset $\Delta_0$'. Fig.~\ref{lab:comparison_nz} shows the resulting difference in the recovered redshift distributions, where the `pessimistic' case of Model 1-4 shows considerable biases that are reduced by S1A2P4 and to better accuracy by S1A5P4. This recovered bias is naturally similar to the fiducial case discussed in the previous section\footnote{Making the ansatz of Model 5 for the photometric sample is very similar to the case studied in the previous section. However as the true scale length model of the spectroscopic sample is Model~1, and not Model~5 as in the previous section, the data covariance matrix is slightly different.}.

Fig.~\ref{lab:model_comparison} compares the relative error in the lensing convergence power spectrum for the different model configurations with the statistical error budget to be expected for the lensing survey defined in Tab.~\ref{tab:cosmo_param}. The corresponding numbers are listed in Tab.~\ref{tab:model_comparison}. We see that model S1A1P4 has the largest error in the convergence power spectrum and is clearly rejected by the DIC criterion. However even though S1A2P4 yields a systematic bias that is larger than the statistical error budget, it has a similar DIC as the much better-performing model S1A5P4. This is a result of the degeneracy between different scale length models with the redshift distribution of the photometric sample.
To highlight this, we impose a prior on the redshift distribution bin heights as described in \S \ref{subsec:prior}. The corresponding results for models denoted as `Nz prior' are shown in Fig.~\ref{lab:model_comparison} and Fig.~\ref{lab:comparison_nz}. We see that the inclusion of a prior improves S1A2P4 and S1A5P4 in terms of the relative error in the convergence power spectrum. Most importantly however, the difference in DIC is much larger compared with the case that does not incorporate prior information. The worse S1A2P4 model is now more clearly rejected by the DIC criterion. We note that imposing a prior on the scale length offset $\Delta_0$ was not sufficient in our experiments to produce this effect. This is sensible, as the redshift distribution in our setup contains many more degrees of freedom compared with the parametrization of the scale length model. This indicates that combining the clustering measurement with external redshift constraints from e.g.\ template fitting will be necessary to break the degeneracy between our uncertainty in the redshift-dependent scale factor, i.e.\ the redshift-dependent galaxy-dark matter bias, and the redshift distribution of the photometric sample.

\section{Summary and Conclusions}
\label{sec:summary_and_conclus}
We have presented a hierarchical logistic Gaussian process model to estimate photometric redshift distributions from cross-correlations between the photometric sample and spatially overlapping spectroscopic samples. Using a simulated data vector and a covariance matrix that mimics a current photometric survey, we demonstrated that this model is able to accurately estimate photometric redshift distributions in a fully Bayesian manner. More specifically, we tested the robustness of our approach to biases in the redshift-dependent galaxy-dark matter bias model. For this we use several published scale length model measurements from the Illustris simulation in our likelihood simulation. We then investigated several scenarios that range from small systematic biases in our parametrization of the redshift-dependent scale length, to more extreme cases where the assumed scale length model does not capture the true underlying function. In our model, we match the scale length model in the spectroscopic sample to the scale length of the photometric sample. This is a good approximation, if the spectroscopic and photometric samples have a roughly similar galaxy population as measured by e.g.\ their stellar mass. A test case that would mimic this situation selects a spectroscopic and photometric sample in stellar mass bins from $8.5 < \log{M_{*} [h^{-2} M_{\odot}])} < 9.0$ and $9.0 < \log{M_{*} [h^{-2} M_{\odot}])} < 9.5$ respectively. For these galaxy samples, the scale length parametrization had a low systematic bias and the recovered redshift distribution posteriors were consistent with the truth within the error bars.

To probe the robustness of our redshift inference we then tested a spectroscopic scale length model that corresponds to a galaxy sample in a much higher stellar mass bin of $10.5 < \log{M_{*} [h^{-2} M_{\odot}]} < 11.0$, where we expect our parametrization to under-perform. For this more pessimistic case we found indeed that the redshift posterior distribution is significantly biased. Forecasting the systematic biases in the lensing convergence power spectrum, we find that these biases exceed the statistical error budget in the measurement by a factor of 3, assuming a DES Y5-like lensing survey.
In order to detect these biases we propose to evaluate a range of models in a Bayesian model comparison to find the `best fitting' candidate. We tested several different scale length models and found that our model comparison statistic reliably detects models with a large systematic modelling bias.

We finally test how the incorporation of additional information from e.g.\ template-based redshift inference affects the prediction performance and the results of the model comparison. We set a prior on our logistic Gaussian process using a Gaussian approximation to a Bayesian histogram posterior. This mimics a photometric redshift distribution obtained by a strongly idealized photometric redshift code. By sampling our model using this prior information we find that the quality of our redshift inferences is significantly improved. Furthermore we find that the sensitivity of our model comparison statistic to systematic biases in the scale length model is also significantly enhanced.
We therefore conclude that the combination of SED-based redshift estimation and cross-correlation will strongly benefit the accuracy and robustness of the overall redshift inference. In a future publication we will extend this model to include a template fitting likelihood. {\color{black} To improve the sampling efficiency, specifically in the low signal-to-noise tails of the distribution, we will also include a variable selection that reduces the number of free parameters that need to be sampled. More importantly this will enable us to marginalize over cosmological parameters in a joint analysis, for which computationally more expensive likelihoods must be evaluated.} We will also apply our methodology to more realistic spectroscopic calibration samples with a more realistic galaxy type selection.

\section*{Acknowledgements}
MMR is supported by DOE grant DESC0011114 and NSF grant IIS-1563887. RM is partially supported by NSF grant IIS-1563887.




\bibliographystyle{mnras}
\bibliography{bibliography.bib} 








\bsp	
\label{lastpage}
\end{document}